\documentclass[sn-mathphys-num]{sn-jnl}

\usepackage{enumitem}
\usepackage{graphicx}%
\usepackage{multirow}%
\usepackage{amsmath,amssymb,amsfonts}%
\usepackage{amsthm}%
\usepackage{mathrsfs}%
\usepackage[title]{appendix}%
\usepackage{xcolor}%
\usepackage{textcomp}%
\usepackage{manyfoot}%
\usepackage{booktabs}%
\usepackage{algorithm}%
\usepackage{algorithmicx}%
\usepackage{algpseudocode}%
\usepackage{listings}%
\usepackage{subfig}
\usepackage{tikz}
\usepackage{pdflscape}

\theoremstyle{thmstyleone}%
\theoremstyle{thmstyletwo}%

\theoremstyle{thmstylethree}%

\begin{document}

\title[DL-Polycube for high-quality hex mesh generation and volumetric spline construction]{DL-Polycube: Deep learning enhanced polycube method for high-quality hexahedral
  mesh generation and volumetric spline construction}


\author*[1]{Yuxuan Yu}\email{yuyuxuan@dhu.edu.cn}
\author[1]{Yuzhuo Fang}\email{slccovey@outlook.com}
\author[2]{Hua Tong}\email{huat2@andrew.cmu.edu}
\author*[2]{Yongjie Jessica Zhang}\email{jessicaz@andrew.cmu.edu}


\affil*[1]{\orgdiv{Institute of Artificial Intelligence}, \orgname{Donghua University}, \orgaddress{\street{2999 North Renmin Road}, \city{Shanghai}, \postcode{201620}, \country{China}}}

\affil[2]{\orgdiv{Department of Mechanical Engineering}, \orgname{Carnegie Mellon University}, \orgaddress{\street{5000 Forbes Ave}, \city{Pittsburgh}, \postcode{15213}, \state{PA}, \country{USA}}}


\abstract{In this paper, we present a novel algorithm that integrates deep
  learning with the polycube method (DL-Polycube) to generate high-quality
  hexahedral (hex) meshes, which are then used to construct volumetric splines
  for isogeometric analysis. Our DL-Polycube algorithm begins by establishing a
  connection between surface triangular meshes and polycube structures. We
  employ deep neural network to classify surface triangular meshes into their
  corresponding polycube structures. Following this, we combine the acquired
  polycube structural information with unsupervised learning to perform surface
  segmentation of triangular meshes. This step addresses the issue of
  segmentation not corresponding to a polycube while reducing manual
  intervention. Quality hex meshes are then generated from the polycube
  structures, with employing octree subdivision, parametric mapping and quality
  improvement techniques. The incorporation of deep learning for creating
  polycube structures, combined with unsupervised learning for segmentation of
  surface triangular meshes, substantially accelerates hex mesh
  generation. Finally, truncated hierarchical B-splines are constructed on the
  generated hex meshes. We extract trivariate B\'{e}zier elements from these
  splines and apply them directly in isogeometric analysis. We offer several
  examples to demonstrate the robustness of our DL-Polycube algorithm.}


\keywords{polycube, graph convolutional network, hexahedral mesh generation,
  volumetric spline construction, isogeometric analysis}



\maketitle

\section{Introduction}\label{sec:introduction}


Isogeometric analysis (IGA) is an analysis method designed to unify finite
element analysis (FEA) and computer-aided design (CAD) by using the same basis
functions for both geometrical and simulation
representation~\cite{cottrell_isogeometric_2009}. Introduced by T.J.R. Hughes in
2005~\cite{Hughes05a}, IGA has evolved significantly over the past two
decades. Despite these advancements, constructing volumetric parameterization
from surfaces remains a challenging task. Most CAD software represents geometry
using boundary representation (B-Rep) models, typically expressed mathematically
through NURBS (Non-Uniform Rational B-Splines). These B-Rep models are composed
of multiple B-spline or NURBS patches.  Consequently, IGA initially focused on
B-Rep models~\cite{Hughes05a}. In 2003, T.W. Sederberg introduced
T-splines~\cite{ref:sederberg03}, which allow for T-junctions in quadrilateral
control meshes, enabling local refinement. However, traditional T-splines lack
linear independence and partition of unity properties, making them unsuitable
for IGA and design. Recent research has developed analysis-suitable T-splines
for IGA~\cite{li_analysis-suitable_2014}, particularly for shell
structures~\cite{wei_analysis-suitable_2022}. Although B-Rep model
parameterization is often used in IGA, three-dimensional (3D) solid models often
have advantages over B-Rep models which discard internal geometric
information. IGA often requires volumetric representation to account for
internal material structures and densities. Therefore, constructing volumetric
spline parameterization models suitable for IGA has been a persistent challenge
in applying IGA to 3D solid models.

Research on constructing volumetric parameterization for IGA can be categorized
into two main approaches based on the input: constructive solid
geometry (CSG)~\cite{zuo_isogeometric_2015} and
B-Rep~\cite{li_generalized_2010,zhang_solid_2012}. However, CSG-based models
pose difficulties for IGA due to the presence of trimming surfaces. B-Rep models
require generating control meshes and constructing volumetric spline basis
functions. In FEA, B-Rep models can be discretized into tetrahedral or
hexahedral (hex) meshes. While tetrahedral mesh generation has multiple automatic
strategies and is widely used in industry, hex meshes are preferred for
their advantages: fewer elements for the same
accuracy~\cite{benzley1995comparison}, avoidance of tetrahedral
locking~\cite{francu2021locking}, and better suitability for tensor-product
spline construction.

Despite ongoing research in hex mesh
generation~\cite{pietroni_hex-mesh_2023,ref:zhangbook,zhang2013challenges},
generating high-quality hex meshes for complex B-Rep models remains challenging.
Methods like indirect methods~\cite{E96}, sweeping
methods~\cite{Zhang20072943,yu2019anatomically}, grid-based
methods~\cite{qian2012automatic,schneiders1996grid,qian2010quality}, polycube
methods~\cite{Tarini2004,Wang07polycubesplines,LeiLiu2012a,HZ2015CMAME,yu2020hexgen},
and vector field-based methods~\cite{nieser2011cubecover,Li2012AMU} have been
explored. However, not all these methods are suitable for IGA. Hex meshes often
involve extraordinary points (an extraordinary point has other than four faces
adjacent to it) or extraordinary edges (an extraordinary edge is an interior
edge shared by other than four hex elements), which can complicate spline
construction and parameterization. When extraordinary points or extraordinary
edges are involved, achieving optimal convergence rates is an challenging
problem in IGA, even smooth basis functions are defined around them. To minimize
the number of extraordinary points and extraordinary edges, sweeping and
polycube methods are preferred for generating hex control meshes for
IGA. Sweeping methods produce high-quality meshes by scanning from source to
target surfaces, but their applicability is limited to specific models where the
source and target surfaces have similar topology. Polycube methods use polycubes
as parameter spaces to generate hex meshes through parametric mapping. The
concept of polycubes was first introduced as a texture mapping technique for
general meshes~\cite{Tarini2004}. Lin \textit{et al}.~\cite{LinJ2008} proposed
an automated method for constructing polycubes; however, this method is not
suitable for models with complex topology and geometry. Due to the polycube
method's capability of controlling the location of extraordinary points and
edges, and because the polycube structure and parametric mapping can be used
together to generate 3D T-spline models~\cite{zhang_solid_2012}, the polycube
method has been employed to generate hex control meshes. Hu \textit{et
  al}.~\cite{HZ2015CMAME,HZL2016} used centroidal Voronoi tessellation to
segment surfaces and generated hex control meshes using the polycube method. Guo
\textit{et al}.~\cite{guo_cut-enhanced_2020} introduced extraordinary edges
within the polycube to improve mesh quality. Li \textit{et al}.~\cite{LiBo2012}
proposed the generalized polycube method, significantly improving the
adaptability of the polycube method to high-genus B-Rep models. Yu \textit{et
  al}.~\cite{yu2020hexgen} used CVT to segment surfaces and generated
generalized polycubes with internal extraordinary edges, subsequently generating
hex control meshes through the generalized polycube method.

Constructing splines on hex control meshes is another challenge. Various
spline techniques like NURBS~\cite{Zhang20072943},
T-splines~\cite{Wang07polycubesplines,zhang_solid_2012,LeiLiu2012a}, and
TH-splines~\cite{wei17a} have been developed for volumetric parameterization. It
involves achieving optimal convergence rates and minimizing geometric
errors. Optimal convergence rates are related to the extraordinary points and
edges of the control mesh. For geometric errors, global or local refinement
combined with spline fitting can be employed. Compared to globally subdividing
the entire space, local refinement can accurately fit surface geometry using
fewer control points. Xu \textit{et al}.~\cite{xu2019efficient} optimized the positions
of control points in regions with significant geometric features and errors
through r-adaptive framework. Li \textit{et al}.~\cite{LiBo2012} used the local
refinement capability of T-splines to reduce geometric errors when fitting
surface models. However, for B-Rep models with complex surface features, regions
with significant geometric errors usually occur around the geometric features. If
the control mesh does not preserve these features, it can indirectly affect the
magnitude of geometric errors. Therefore, when global or local refinement cannot
significantly reduce geometric errors, further processing of the control mesh is
necessary, which includes adding feature information to the hex control
mesh.

To address these challenges, we developed HexGen and
Hex2Spline~\cite{yu2020hexgen} in our previous research. HexGen generates hex
meshes, while Hex2Spline uses them as the input control meshes to construct
volumetric truncated hierarchical splines. Through B\'{e}zier extraction,
Hex2Spline transfers spline information to ANSYS-DYNA for IGA. Our results
demonstrate the algorithm's capability to generate analysis-suitable volumetric
parameterization models for IGA. However, the process involves significant
manual work in the input file preparation, which includes surface segmentation
and polycube construction. The automation of surface segmentation and polycube
construction was also discussed in a 2023 survey
paper~\cite{pietroni_hex-mesh_2023}, where it was defined as labeling. In this
context, the polycube structure can be defined by assigning a label to each
surface triangular element of the input mesh. Early labeling methods used the
six principal axes ($ \pm X$, $ \pm Y$, $ \pm Z$). The first issue is that this
approach is suitable for models where any face is roughly perpendicular to one
of the six normal directions. However, forcing a regular structured polycube to
fit some complex geometries results in poor quality meshes. Another issue is
that different approaches produce different labelings, such as the centroidal
Voronoi tessellation method~\cite{HZ2015CMAME}, local
approaches~\cite{Gregson2011}, and incremental
approaches~\cite{mandad2022intrinsic}. In addition, not all labelings result in
a valid polycube. Some
researchers~\cite{Gregson2011,livesu2013polycut,eppstein2010steinitz} have
proposed adjustment methods, but these methods are neither robust nor
straightforward~\cite{pietroni_hex-mesh_2023}.

In this paper, we focus on replacing the manual surface segmentation process and
polycube construction with intelligent, automated methods. To reduce manual
labor, we integrate deep learning into the polycube method, developing the deep
learning-enhanced polycube (DL-Polycube) algorithm. This method predicts the
polycube structure of input geometries and segments geometry surfaces to match
the polycube structure, automating the surface segmentation and polycube
construction process. The DL-Polycube algorithm significantly enhances the
efficiency of our previously developed HexGen and Hex2Spline software packages
by minimizing the manual effort required for surface segmentation and polycube
construction. In summary, there are three main contributions in this paper.

\begin{enumerate}
\item Unlike the traditional sequence of generating hex meshes based on
  polycubes, the DL-Polycube algorithm employs machine learning to automatically
  construct polycube structures from input geometries and then perform surface
  segmentation based on these polycube structures. This approach aligns more
  closely with domain knowledge or human intuition. A skilled mesh generation
  engineer typically spends time optimally deciding how to partition complex
  geometries into simpler regions before manually generating hex
  meshes. They then create quadrilateral surface meshes on each partitioned
  block and use mapping or sweeping methods to generate hex meshes within
  the volumes. Our method approximates this process using machine learning.
\item The DL-Polycube algorithm demonstrates its capability in rapidly
  predicting the corresponding polycube structure of geometric models and
  segmenting the surface of CAD geometry to achieve a one-to-one correspondence
  with the surface of the polycube structure. By integrating deep learning with
  the polycube method, we automate the process of converting CAD geometries into
  high-quality hex meshes and volumetric spline models. The entire prediction
  step completes in one second for a single model. Our approach demonstrates
  that machine learning can effectively automate and optimize the mesh
  generation process.
\item The DL-Polycube algorithm enhances the efficiency of our previously
  developed HexGen and Hex2Spline software packages. Although these packages can
  handle complex models, they require suitable input files, which involve manual
  work for surface segmentation and polycube structure construction. By
  integrating deep learning, we reduce this manual effort by replacing the
  manual surface segmentation process and polycube construction with
  intelligent, automated methods. This significantly reduces the learning curve
  for the software.
\end{enumerate}

The remainder of this paper is organized as follows. Section 2 provides an
overview of the pipeline, including the design of the DL-Polycube
algorithm. Section 3 details the implementation of the DL-polycube method,
including dataset generation and curation, and the machine learning
model. Section 4 discusses machine learning-driven polycube-based segmentation
and path optimization for zigzag issues. Section 5 covers high-quality hex mesh
generation and volumetric spline construction. Section 6 presents examples to
demonstrate the feasibility and efficiency of the algorithm in generating hex
meshes and splines. Finally, Section 7 concludes and suggests future work.

\section{Overview of the pipeline}

\subsection{Pipeline design}

\begin{figure}[!htb]
      \centering
  \begin{tikzpicture}
    \node[anchor=south west,inner sep=0] (image) at (0,0) {\includegraphics[width=\linewidth]{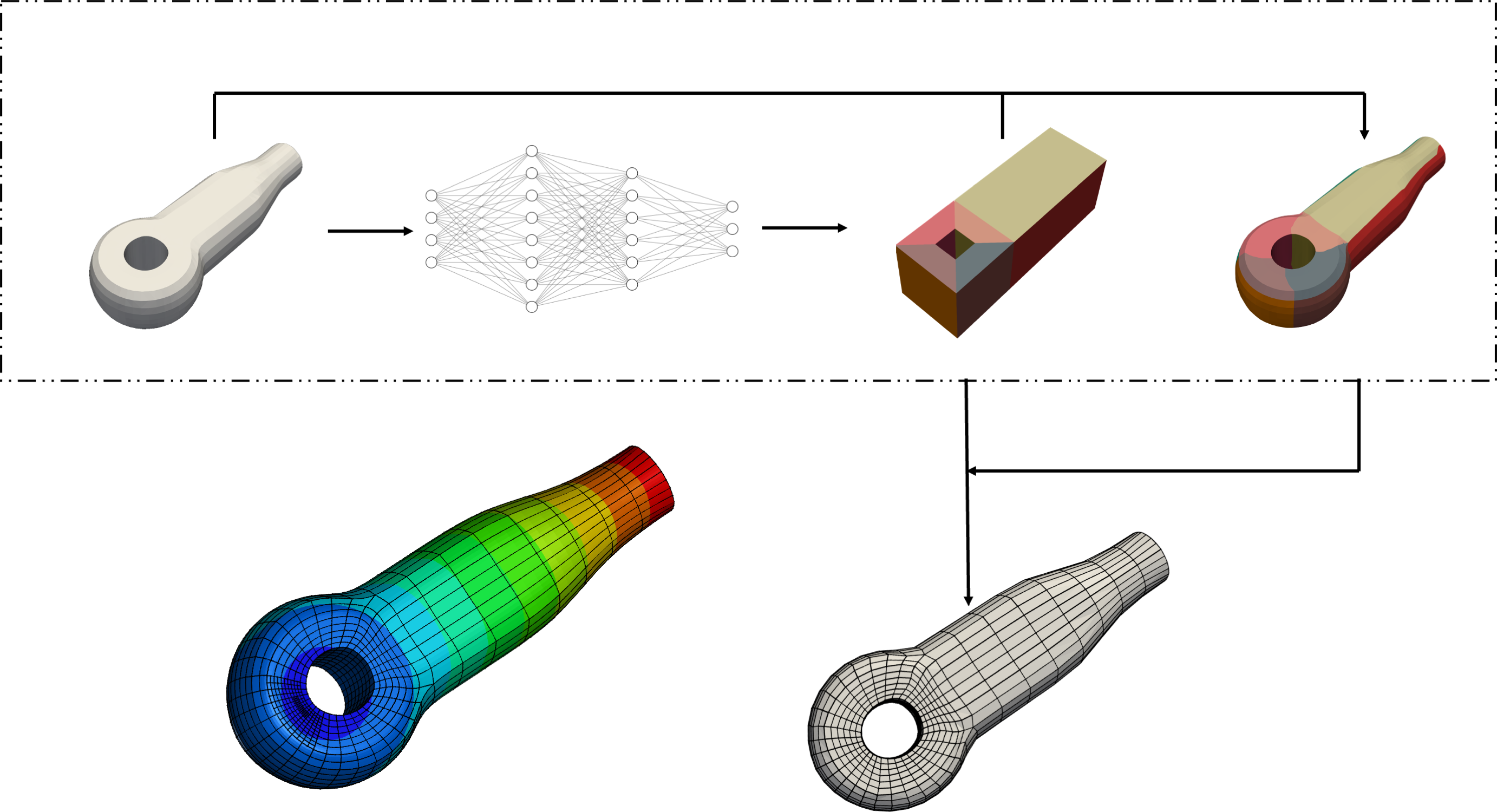}};
    \begin{scope}[x={(image.south east)},y={(image.north west)}]
      \node at (0.11,0.56) {(a)};
      \node at (0.38,0.56) {(b)};
      \node at (0.64,0.56) {(c)};
      \node at (0.88,0.56) {(d)};
      \node at (0.26,-0.015) {(f)};
      \node at (0.66,-0.015) {(e)};
      \node at (0.1,0.96) {\textbf{DL-Polycube}};
      \node at (0.5,0.92) {K-means};
      \node at (0.78,0.45) {Parametric mapping};
      \node at (0.78,0.50) {Octree subdivision};
      \node at (0.78,0.38) {Quality improvement};
    \end{scope}
  \end{tikzpicture}
  \caption{\label{fig:Pipeline_Structure}{The DL-Polycube pipeline using deep
      learning and the polycube method. (a) The CAD geometry and converting it
      into a triangular mesh; (b) a pretrained deep learning model to generate a
      polycube structure from the triangular mesh; (c) the polycube structure
      predicted by the deep learning model; (d) surface segmentation using the
      polycube structure information with K-means segmentation; (e) the polycube
      structure and surface segmentation are used to create an all-hex mesh
      through octree subdivision, parametric mapping and quality improvement
      techniques; and (f) volumetric spline with IGA simulation results using
      ANSYS-DYNA.}}
\end{figure}
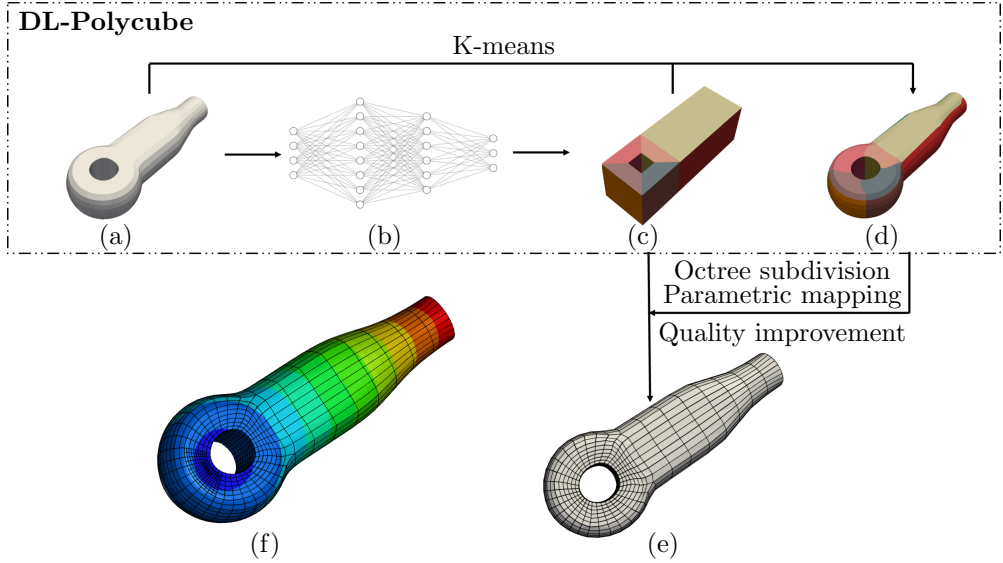

We integrate deep learning with the polycube method to convert CAD geometries
into volumetric spline models. As shown in Fig.~\ref{fig:Pipeline_Structure},
our DL-Polycube pipeline begins with the conversion of the CAD geometry into a
triangular mesh. Subsequently, a pretrained deep learning model is used to
generate a polycube structure. This structure serves as the foundation for
generating all-hex meshes through parametric mapping~\cite{floater1997} and
octree subdivision techniques~\cite{zhang_solid_2012}.  To ensure the all-hex
mesh meets the quality requirements necessary for IGA, we evaluate the all-hex
mesh and employ several mesh quality improvement
techniques--pillowing~\cite{YZhang2009c}, smoothing, and
optimization~\cite{qian2012automatic, tong_hybridoctree_hex_2024}--as needed.
Upon achieving a good quality hex mesh, the volumetric spline model is
constructed from the hex mesh using TH-spline3D with local refinement. Then the
B\'ezier information is extracted to perform IGA in ANSYS-DYNA.

\subsection{DL-Polycube algorithm}

\begin{figure}[!htb]
  \centering
  \begin{tikzpicture}
    \node[anchor=south west,inner sep=0] (image) at (0,0) {\includegraphics[width=\linewidth]{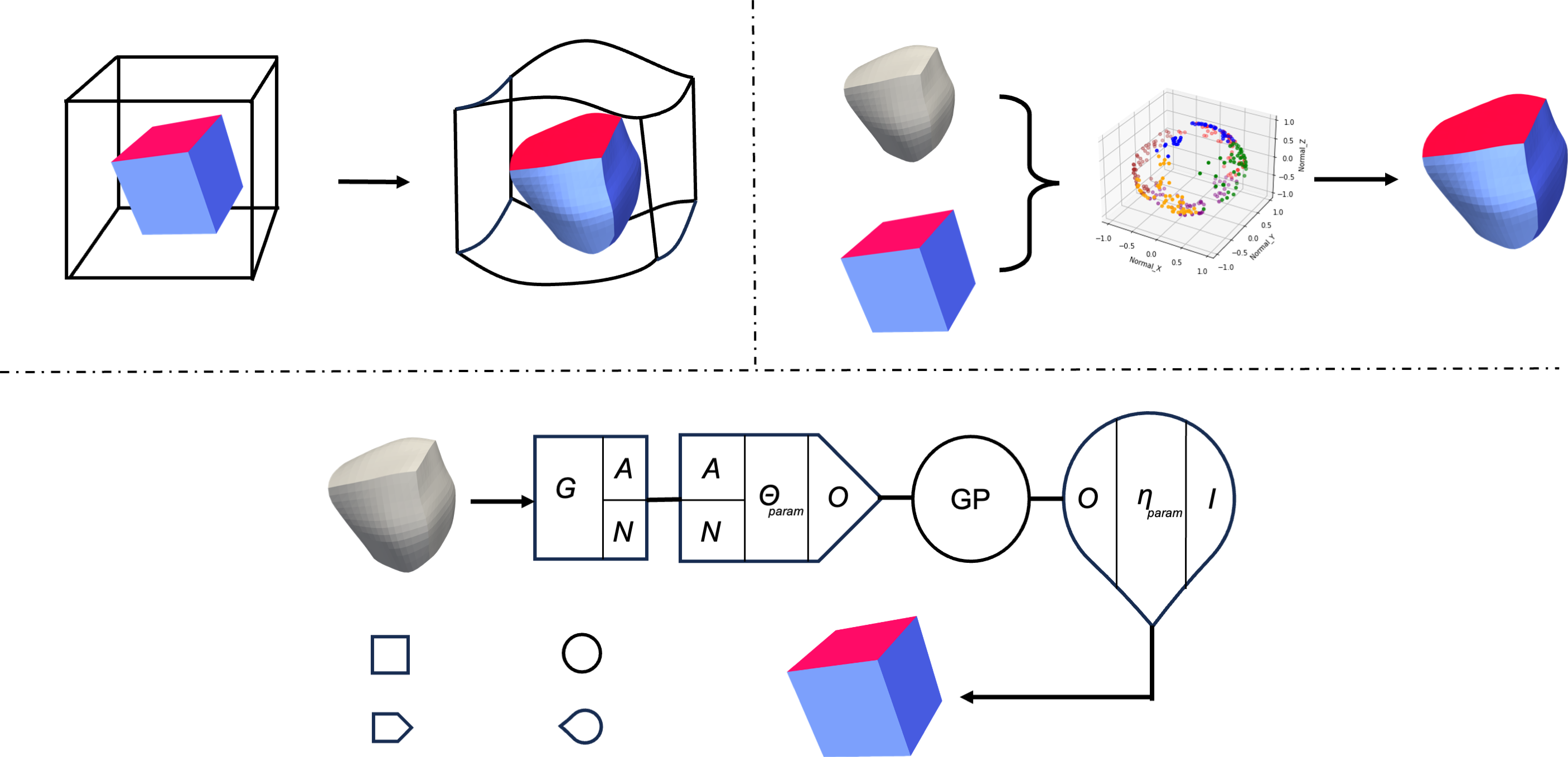}};
    \begin{scope}[x={(image.south east)},y={(image.north west)}]
      \node at (0.05,0.98) {\textbf{Step 1}};
      \node at (0.54,0.98) {\textbf{Step 3}};
      \node at (0.22,0.55) {Dataset};
      \node at (0.74,0.55) {K-means};
      \node at (0.24,0.46) {\textbf{Step 2}};
      \node at (0.5,-0.08) {GCN-Polycube};
      \node at (0.30,0.135) {\small Graph};
      \node at (0.44,0.135) [align=center] {\small Global\\[-0.3em]pooling};
      \node at (0.30,0.025) [align=center] {\small Gconv\\[-0.3em]layers};
      \node at (0.44,0.025) [align=center] {\small Linear\\[-0.3em]layers};
    \end{scope}
  \end{tikzpicture}
  \caption{\label{fig:ML_architecture} Overview of the DL-Polycube
    algorithm. Step 1: Training dataset generation and curation: A cage is
    created around the target mesh to generate a diverse set of models with
    random transformations and deformations. Step 2: GCN-Polycube recognition:
    The GCN-Polycube classification enables the automatic conversion of complex
    geometries into polycube structures. Step 3: K-means surface segmentation:
    Segment the surface of CAD geometry to ensure a one-to-one correspondence
    with the surface of the polycube structure.}
\end{figure}

The DL-Polycube algorithm includes three steps, which are training dataset
generation and curation, graph convolutional network polycube (GCN-polycube)
recognition, and K-means surface segmentation (see
Fig.~\ref{fig:ML_architecture}). This DL-Polycube method leverages the
efficiency of machine learning to handle a wide range of CAD geometries,
replacing the manual surface segmentation and polycube construction [22] with
intelligent, automated methods. The success of this algorithm is based on the
following assumption: \textit{although there are inconsistencies in the
  geometry models, the underlying polycube structure may remain topologically
  consistent}. By utilizing a pretrained machine learning model, the algorithm
can identify the most suitable polycube for unfamiliar geometries, thereby
automating and optimizing the process. Consequently, this leads to a powerful
tool for generating high-quality hex meshes and constructing volumetric splines
from CAD geometry.

In the dataset generation and curation step, we create a corresponding cage
around the target mesh that requires deformation (e.g., a cube). The deformation
of the cage influences the inner mesh. Deformations are performed by
randomly selecting points on the cage, stretching them to lengths that
follow a normal distribution. The entire process is conducted in Blender, a free
and open-source 3D creation suite, producing effects similar to classical
free-form deformation~\cite{botsch_polygon_2010,sederberg1986}. This method
enables the generation of a diverse set of models with random transformations
and deformations.

In the GCN-Polycube step, GCN-Polycube classification is used for polycube
structure recognition.  DL-Polycube employs multilayer perceptron (MLP)-based
GCN-Polycube models, where MLPs serve as nonlinear regressors. The use of ReLU
activation functions within these MLPs enhances the network's nonlinear
capabilities to fit complex patterns. This approach not only enables DL-Polycube
to automatically convert complex geometries into polycube structures, which
ensures the generation of high-quality all-hex meshes, but also ensures rapid
computations faster than traditional polycube-based
methods~\cite{HZ2015CMAME,HZL2016,Liu2015,guo_cut-enhanced_2020,LeiLiu2012a,Wenyan2013c,zhang_solid_2012}.

During the training step, the ML model undergoes updates in polycube structure
recognition and the K-means surface segmentation. The entire process eliminates
the need for manual intervention, focusing instead on the efficient application
of the trained model to new CAD geometries.

Regarding computational performance, generating and curating the dataset takes
approximately 10 hours to produce 9,900 random meshes. Training the
classification and regression networks for GCN-Polycube classification require
an additional hour. During the prediction step, GCN-Polycube classification is
completed in one second.

The DL-Polycube algorithm demonstrates its capability in rapidly predicting the
corresponding polycube structure of geometric models and segmenting the surface
of CAD geometry to achieve a one-to-one correspondence with the surface of the
polycube structure. Consequently, this tool enables designers to convert
boundary representation geometry into its corresponding polycube structure and
facilitates automatic polycube-based hex mesh generation, as well as the
construction of volumetric splines from CAD geometry. While DL-Polycube
currently lacks parallel processing capabilities beyond the inherent parallelism
of neural networks, its efficient computational framework makes it an important
tool for creating polycube structures and generating hex meshes.

\section{Implementation details of DL-Polycube}

In the initial phase of our pipeline (see Fig.~\ref{fig:Pipeline_Structure}), we
employ a pretrained model to convert CAD geometries into their corresponding
polycube structures. Then, with the generated polycube structures and CAD
models, we segment them into patches that have a one-to-one correspondence
between the surface of the CAD and polycube structures. This process involves
three key steps. Firstly, we generate ``pseudo-random data" to establish a
foundational relationship that links random surface triangular meshes with their
corresponding polycube structures. Following this, a deep learning GCN-Polycube
model architecture, is trained using the pseudo-random data and their
corresponding polycube structures from the first step. This step learns the
mapping between triangular meshes and polycube structures. Finally, we use the
K-means clustering method, which is based on spatial attributes such as normal
vectors and centroids obtained based on the polycube structures. This method
segments the surface by classifying each triangular element into one patch. This
entire process aims at automatically achieving surface segmentation and its
corresponding polycube structure.

\subsection{Dataset generation and curation}
\subsubsection{Procedural geometric modeling}
\label{sec:proc-geom-model}
\begin{figure}[!htb]
  \centering
  \begin{tikzpicture}
    \node[anchor=south west,inner sep=0] (image) at (0,0) {\includegraphics[width=\linewidth]{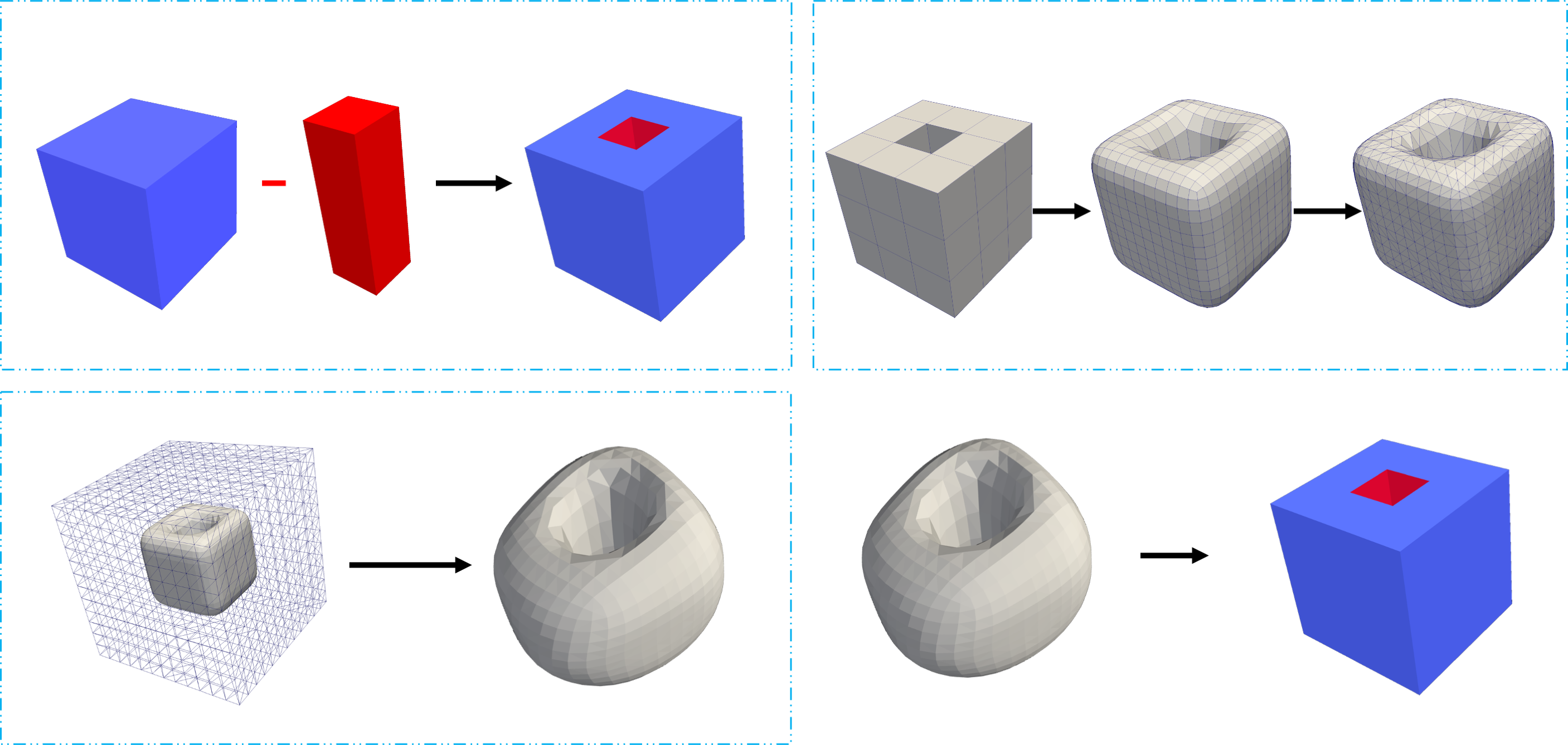}};
    \begin{scope}[x={(image.south east)},y={(image.north west)}]
      \node at (0.05,0.96) {\textbf{Step 1}};
      \node at (0.58,0.96) {\textbf{Step 2}};
      \node at (0.27,0.54) {\small Boolean operation};
      \node at (0.68,0.56) [align=center] {\small Surface\\[-0.3em]subdivision};
      \node at (0.88,0.54) {\small Triangulation};
      \node at (0.05,0.44) {\textbf{Step 3}};
      \node at (0.25,0.03) {\small Deformation};
      \node at (0.64,0.03) {\small Input};
      \node at (0.9,0.03) {\small Output};
    \end{scope}
  \end{tikzpicture}
  \caption{\label{fig:pgmp} Procedural geometric modeling process of
    GCN-Polycube model architecture. Step 1: Initial polycube structure created
    using Boolean operations. Step 2: Surface subdivision applied using the
    Catmull-Clark algorithm and triangulation of the subdivided surface. Step 3:
    Free-form deformation applied to the triangulated and subdivided polycube
    structure. These steps generate input and output geometries for deep
    learning model training.  }
\end{figure}
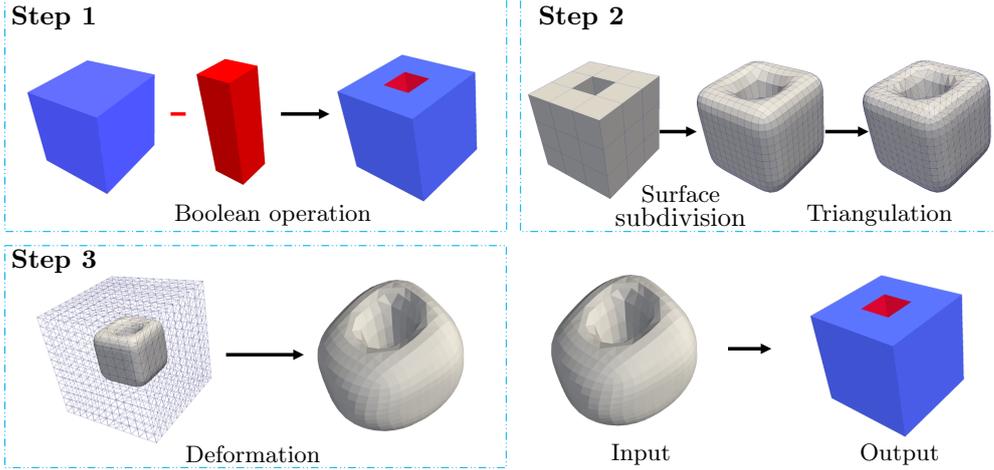

We use the 3D graphics software Blender, along with its built-in Python and
BMesh libraries, to create polycube structures and a wide range of derivative
geometries resulting from their transformations and deformations. These
derivative geometries and their corresponding polycube structures are used as
inputs and outputs for our deep learning model system. This procedural geometric
model employs a multi-step strategy to compute the polycube structures and their
derivative geometries. The first step involves using Boolean operations to
generate polycube structure, as illustrated in Fig.~\ref{fig:pgmp}. The second
step involves surface subdivision using the Catmull-Clark algorithm, followed by
triangulation of the generated polycube structures. The third step involves
free-form deformation. This procedural geometric modeling process is configured
to produce polycube structures and subsequent derivative geometries. Starting
with a polycube structure in the first step, the second and third steps generate
900 derivative geometries corresponding to each initial polycube
structure. Thus, the entire procedural geometric modeling process generates the
input and output data for training.

Finally, we employ procedural geometric modeling to obtain eleven types of
polycube structures and their derivative geometries. These polycube structures
correspond to configurations of a single cube, a genus-one cube, and
combinations of these two primitive geometries. For each type of polycube
structure, we generate 900 corresponding derivative geometries. Note that we
only generate eleven polycube structures as a proof of concept. One can generate
more polycube structures to make the pipeline work for other types of
geometries.

\subsubsection{Data generation}

We employ a parametric script to create various polycube designs, often starting
with basic shapes like cuboids. This script begins by using Boolean operation to
combine basic shapes into a polycube structure and then alters its transformation
and deformation behaviors through procedural geometric modeling, as mentioned in
the previous section. Consequently, the resulting geometry exhibits diverse
shapes and transformation and deformation behaviors while maintaining topological
consistency. It is important to note that the variability of the design
parameters constrains the generalizability of the machine learning model. If the
geometry model encounters design parameters outside its trained range, it is
likely to yield less accurate results. Therefore, these factors must be
considered and integrated throughout the entire pipeline when developing a model
for automatically generating hex meshes.

We use a cuboid primitive with a hole (genus-1) as an example to illustrate our
process (see Fig.~\ref{fig:pgmp}). Firstly, we create the cuboid primitive with
a hole using Boolean operation. Then, we refine it through surface subdivision
using the Catmull-Clark algorithm in Blender. Next, we convert it into a
triangular mesh through triangulation and scale it to a specified size. It is
important to note that scaling to a specified size can play an important role in
feature scaling during model training. Subsequently, we surround the cuboid with
a hole with a cage and apply random stretching to achieve random deformations of
the basic mesh.  Finally, these random meshes serve as inputs for training the
neural network in the subsequent reverse process.

\subsubsection{Feature extraction}

The deformed geometries derived from the initial polycube structures are
used for feature extraction to create the training dataset. Given an input
surface triangular mesh $T$ from the derivative geometry, the geometry can
be represented as an abstract graph G = (A, N), where the adjacency matrix
$\mathrm{A}=\left\{A_{ij}\right\}_{i=1 \ldots N_n, j=1 \ldots N_n}$ defines the
connectivity between triangular elements, with \(A_{ij}=1\) indicating that two
triangular elements are adjacent. The node feature vectors
$\mathrm{N}=\left\{N_i\right\}_{i=1 \ldots N_n}$ encode the characteristics of
each triangular element. Each node feature vector includes details of the
triangular element, such as the vertices that form it and 3D position
$\mathbf{p}$ of each vertex $v_i \in \mathcal{V}$:
\begin{equation}
  \mathbf{p}_i:=\mathbf{p}\left(v_i\right)=\left(\begin{array}{l}
    x\left(v_i\right) \\
    y\left(v_i\right) \\
    z\left(v_i\right)
  \end{array}\right) \in \mathbb{R}^3.
\end{equation}
Besides the physical information, the node feature vectors
$\mathrm{N}=\left\{N_i\right\}_{i=1 \ldots N_n}$ also include additional
attributes such as the normal vector of each face. This dataset, which includes
detailed geometric and topological information, is designed for GCN-Polycube
classification and K-means surface segmentation.

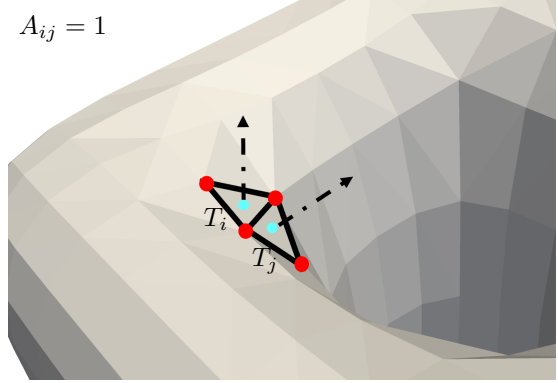
\begin{figure}[!htb]
  \centering
  \begin{tikzpicture}
    \node[anchor=south west,inner sep=0] (image) at (0,0) {\includegraphics[width=0.55\linewidth]{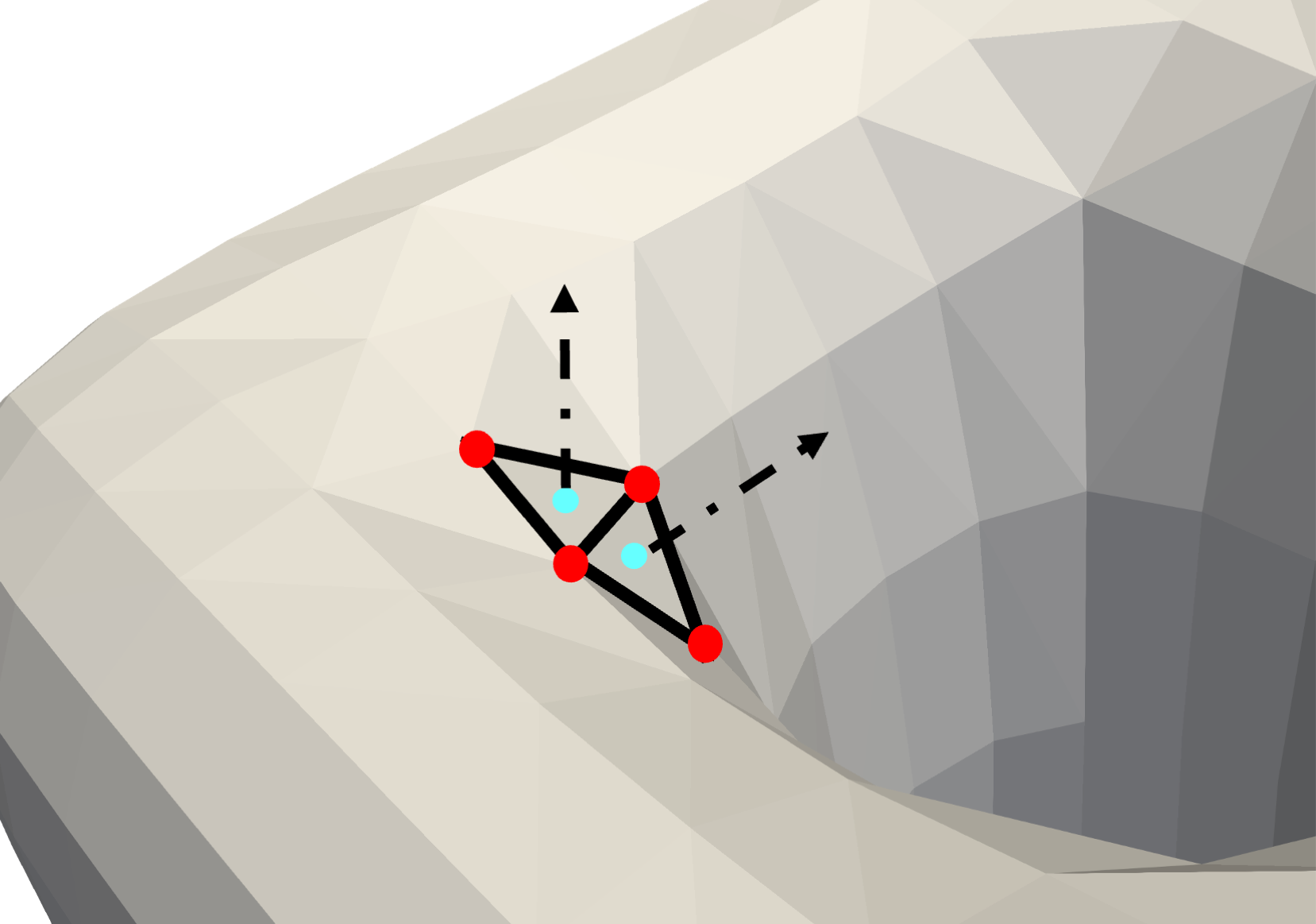}};
    \begin{scope}[x={(image.south east)},y={(image.north west)}]
      \node at (0.1,0.92) {$A_{ij}=1$};
      \node at (0.38,0.42) {$T_i$};
      \node at (0.47,0.31) {$T_j$};
    \end{scope}
  \end{tikzpicture}
  \caption{\label{fig:05_feature} Feature extraction and data integration
    process for GCN-Polycube classification and K-means surface segmentation.
    \(A_{ij}=1\) indicating that two triangular elements $i$ and $j$ are
    adjacent.  The node feature vectors \(N\) include vertex coordinates (red
    points), centroid coordinates (yellow points), and normal vectors (dashed
    arrows) of each face.}
\end{figure}

With the data generated and features extracted, we begin with data labeling.
All derivative geometries are generated through the polycube structure, which
automatically assigns the types of polycube structures as labels to the existing
data. As a result, these derivative geometries and their corresponding polycube
structures serve as inputs and outputs for our deep learning model.  Next, we
perform data curation. We collect and merge data from multiple sources,
including the normal vector in normal space, the vertex coordinates and the
centroid coordinates of triangular elements in 3D Euclidean space, and the
adjacency matrix of the graph (see Fig.~\ref{fig:05_feature}). We use these data
as features. These features include various geometric and graph information to
help our models learn and predict more effectively.

\subsection{Machine learning model}
\label{sec:reverse-process}

The DL-Polycube algorithm aims to automatically generate polycube structures and
ensure a one-to-one correspondence between the surface of CAD geometry and
polycube structures. To achieve this, we will first use GCN-Polycube to extract
the polycube structure of the CAD geometry. Based on the algorithm, the possible
corresponding polycube structures will be given. Then, we will use K-means
segmentation to perform surface segmentation.

\subsubsection{GCN-Polycube model architecture}

\begin{figure}[!htb]
  \centering
  \begin{tikzpicture}
    \node[anchor=south west,inner sep=0] (image) at (0,0) {\includegraphics[width=\linewidth]{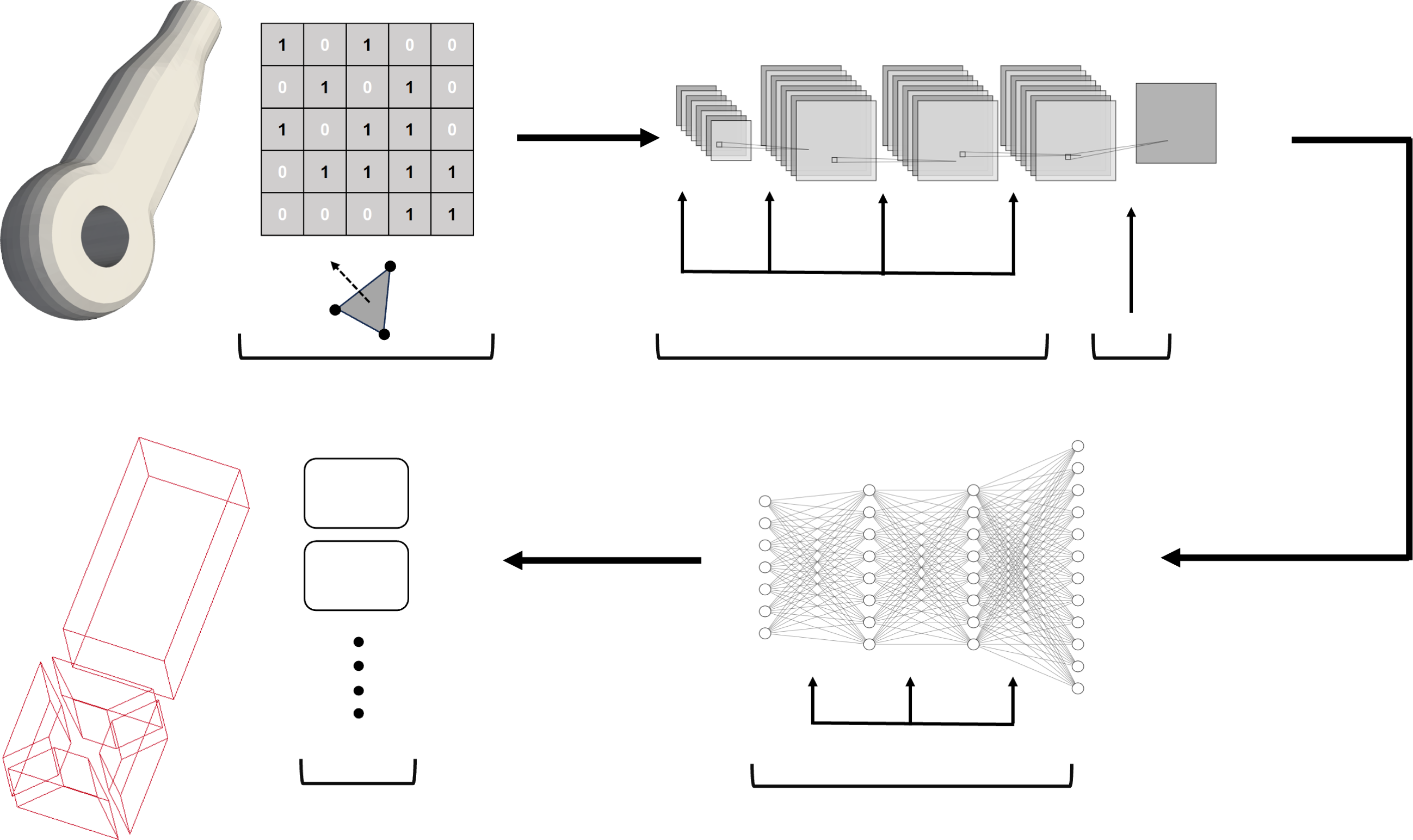}};
    \begin{scope}[x={(image.south east)},y={(image.north west)}]
      \node at (0.26,0.52) {(a)};
      \node at (0.61,0.52) {(b)};
      \node at (0.8,0.52) {(c)};
      \node at (0.255,0.015) {(e)};
      \node at (0.645,0.015) {(d)};
      \node at (0.255,0.415) {\small Type1};
      \node at (0.255,0.315) {\small Type2};
      \node at (0.485,0.64) {\tiny{$12\mkern-6mu\times\mkern-6mu128$}};
      \node at (0.55,0.64) {\tiny{$128\mkern-6mu\times\mkern-6mu256$}};
      \node at (0.625,0.64) {\tiny{$256\mkern-6mu\times\mkern-6mu256$}};
      \node at (0.715,0.64) {\tiny{$256\mkern-6mu\times\mkern-6mu256$}};
      \node at (0.25,0.11) {\tiny{$11\mkern-6mu\times\mkern-6mu1$}};
      \node at (0.57,0.11) {\tiny{$128\mkern-6mu\times\mkern-6mu11$}};
      \node at (0.64,0.11) {\tiny{$128\mkern-6mu\times\mkern-6mu128$}};
      \node at (0.715,0.11) {\tiny{$256\mkern-6mu\times\mkern-6mu128$}};
    \end{scope}
  \end{tikzpicture}
  \caption{\label{fig:hierarch_GC} Hierarchical overview of the GCN-Polycube
    model architecture. (a) The input features from the CAD geometry. (b) The
    four GCN-Polycube layers capture local and global geometric structures. (c)
    The pooling layer aggregates node-level information into graph-level
    representations. (d) The fully connected layers map the features of complex
    topological structures to the target space, i.e., the polycube
    structure. (e) The output predicts the suggest corresponding polycube
    structures. }
  \end{figure}
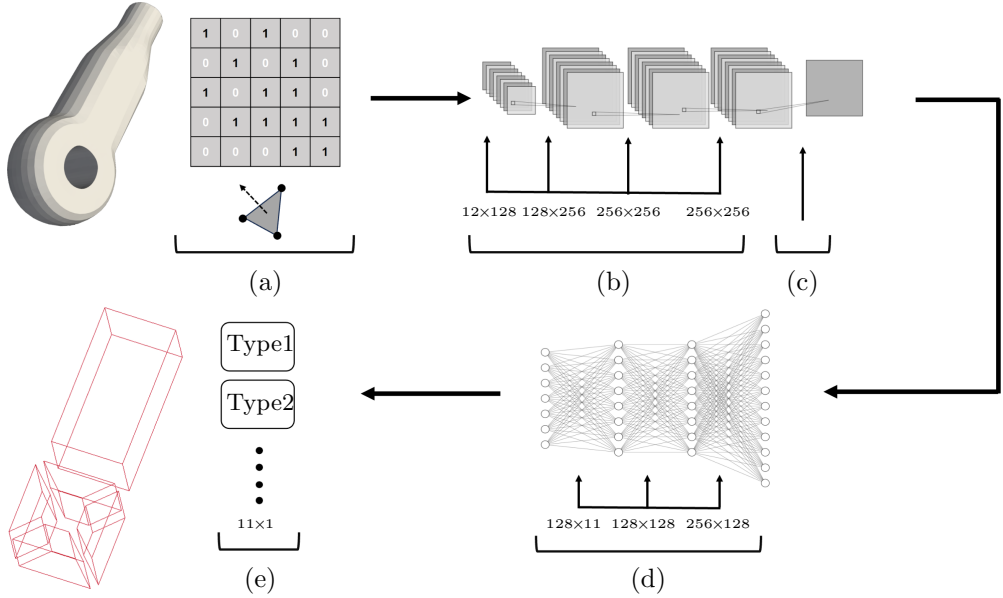

Fig.~\ref{fig:hierarch_GC} provides a hierarchical overview of the GCN-Polycube model
architecture. The architecture is composed of multiple layers designed to
capture and process the spatial relationships and features of the deformed
geometries derived from the initial polycube structures. The initial layer takes
the graphical representation of the CAD geometry as input, transforming it into
a feature-enhanced representation through convolutional operations that consider
both node features and their connections. Subsequent layers in the GCN-Polycube further
refine these features by aggregating information from neighboring nodes,
capturing the local and global geometric structures. This hierarchical
processing allows the model to understand complex spatial dependencies and
interactions within the CAD geometry. The output of the GCN-Polycube is then used to suggest
possible corresponding polycube structures, ensuring a one-to-one correspondence
between the surface of the CAD and polycube models. A layer of a GCN-Polycube can be described by the
following formula~\cite{kipf2016semi}:
\begin{equation}
  F^{(l+1)} = H \left( \tilde{D}^{-1/2} \tilde{A} \tilde{D}^{-1/2}
    F^{(l)} W^{(l)} \right),
  \label{eq:gcn}
\end{equation}
where $F^{(l)}$ represents the feature matrix at layer $l$, $\tilde{A}=A+I$ is
the adjacency matrix of the graph with added self-loops, $\tilde{D}$ is the
degree matrix of $\tilde{A}$, $W^{(l)}$ is the trainable weight matrix of layer
$l$, and $H$ is the activation function ReLU.  Our neural network consists of
those four GCN-Polycube layers followed by a global average pooling, concluding
with a fully connected layer to produce the output. The Algorithm 1 is a snippet
of the forward computation of a GCN-Polycube Layer.

\begin{algorithm}[H]
  \caption{Forward computation of GCN-Polycube layer}
  \begin{algorithmic}[1]
    \State \textbf{Input:} Feature matrix \( F^{(l)} \), Adjacency matrix \( A \), Weight matrix \( W^{(l)} \)
    \State \textbf{Output:} Updated feature matrix \( F^{(l+1)} \)

    \State Add self-loops to adjacency matrix of deformed geometries
    derived from the initial polycube structures
    \State Compute degree matrix of deformed geometries
    \State Compute inverse square root of degree matrix
    \State Normalize adjacency matrix
    \State Linear transformation of normalized adjacency matrix
    \State Apply activation function to transformation result
    \State \textbf{return} \( F^{(l+1)} \)
  \end{algorithmic}
\end{algorithm}

\noindent \textbf{\textit{Loss function in GCN-Polycube}}. The loss function
quantifies the difference between the predicted polycube structures and the true
polycube structures (initial polycube structures in
Sec.~\ref{sec:proc-geom-model}). For GCN-Polycube, we use the cross-entropy (CE)
loss, which measures the performance of our GCN-Polycube model. The CE loss is
defined as:
\begin{equation}
  \mathcal{L}_{\text{CE}} = -\frac{1}{N} \sum_{i=1}^{N} \sum_{c=1}^{C} y_{i,c} \log(\hat{y}_{i,c}) + \lambda \sum_{l} \|W^{(l)}\|^2,
\end{equation}
where \(N\) is the number of training data, \(C\) is the number of types of
polycube structure considered in this paper, \(y_{i,c}\) is the binary indicator
indicating whether the true polycube structure \(c\) is the correct
classification for training data \(i\), and \(\hat{y}_{i,c}\) is the predicted
probability that training data \(i\) belongs to true polycube structure
\(c\). \(L_2\) regularization is adopted to prevent overfitting. Here
\(\lambda\) is a regularization parameter that controls the trade-off between
fitting the data and keeping the model weights small, and \(W^{(l)}\) are the
weights of the \(l\)-th layer shown in Equation~\eqref{eq:gcn}.

\subsubsection{K-means model architecture}
\label{sec:mesh-segmentation}

Step 3 in Fig.~\ref{fig:ML_architecture} provides a hierarchical overview of the
model architecture for K-means surface segmentation using polycube structure
information. By using a trained GCN-Polycube model from the previous step, we
can obtain the predicted polycube structure of the CAD geometry. Subsequently,
we use the K-means algorithm to segment the CAD geometry into $k$ clusters in an
unsupervised manner. Traditionally, the number of clusters $k$ in K-means is a
hyperparameter. However, in this context, $k$ is known because it corresponds to
the $k$ faces of the predicted polycube structure. The $k$ initial centroids are
not chosen randomly; instead, they are calculated by taking the mean of the
normal vectors corresponding to the predicted polycube structure. We first
select $k$ initial centroids based on the predicted polycube structure. Each
normal vector of a face of predicted polycube structure is used as an initial
centroid. Given an input surface triangular mesh $T$ from CAD geometry, let the
dataset $X=\left\{x_{T(i)}\right\}_{i=1}^n$ denotes the unit normal vectors
$x_{T(i)}$ set of the triangular mesh in the normal space, where $n$ is the
total number of triangles and $T(i)$ represents the $i^{th}$ triangle in the
physical space. Then we assign each data point $x_{T(i)}$ to the nearest
centroid based on the Euclidean distance in the normal space. This can be
expressed as:
\begin{equation}
  C_j = \left\{ x_{T(i)} : \| x_{T(i)} - \mu_j \|^2 \leq \| x_{T(i)} - \mu_l
    \|^2 \,\  \forall \, l, 1 \leq l \leq k \right\},
  \label{eq:k-means}
\end{equation}
where $C_j$ is the set of points assigned to centroid $\mu_j$. Then we
recalculate the centroids as the mean of all data points $x_{T(i)}$ assigned to each cluster:
\begin{equation}
  \mu_i = \frac{1}{|C_j|} \sum_{x_{T(i)} \in C_j} x_{T(i)},
\end{equation}
where $|C_j|$ is the number of points in cluster $C_j$. Finally, we repeat the
assignment and update steps until the loss function no longer change
significantly ($<$3\%). The loss function is defined as:
\begin{equation}
  \sum_{j=1}^{k} \sum_{x_{T(i)} \in C_j} \| x_{T(i)} - \mu_j \|^2.
\end{equation}
This results in $k$ clusters and surface of geometry is
segmented into $k$ parts corresponding to the polycube structures. Algorithm
2 is a snippet of the K-means Clustering Algorithm combined with polycube
structure information.

\begin{algorithm}[H]
  \caption{K-means clustering algorithm}
  \begin{algorithmic}[1]
    \State \textbf{Input:} Manifold, watertight triangular mesh $T$, number of clusters $k$ obtained from the initial polycube structure,
    data points $x_{T(1)}, x_{T(2)}, \ldots, x_{T(n)}$, the
    normal vector of each surface of polycube structure $\mu_1, \mu_2, \ldots, \mu_k$
    \State \textbf{Output:} Surface segmentation results

    \State Initialize $k$ centroids of data as $\mu_1, \mu_2, \ldots, \mu_k$
    \Repeat
    \State \textbf{Assignment step:}
    \For{each data point $x_{T(i)}$}
    \State Assign $x_{T(i)}$ to the cluster $C_j$ with the nearest centroid
    \EndFor
    \State \textbf{Update step:}
    \For{each cluster $C_j$}
    \State Update the centroid $\mu_j$ to the mean of all points in $C_j$
    \EndFor
    \Until{centroids do not change significantly}
    \State Obtain cluster assignment for each data point and its
    corresponding cluster centroid
    \State Cluster assignments lead to surface segmentation
  \end{algorithmic}
\end{algorithm}

\subsection{Model training in GCN-Polycube and K-means}

\begin{figure}[!htp]
\centering
\begin{tabular}{cc}
  \includegraphics[width=0.5\linewidth]{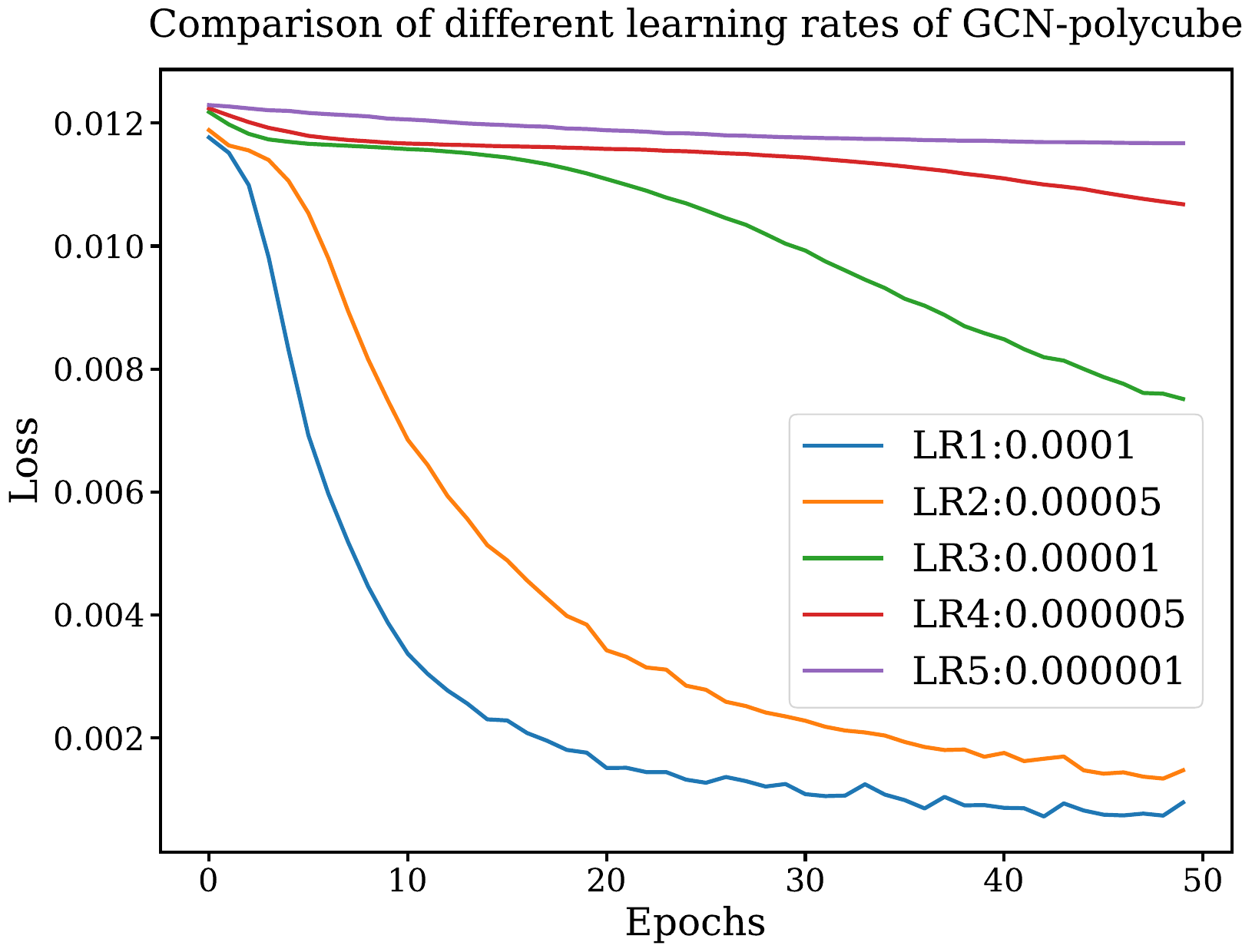}
  &\includegraphics[width=0.485\linewidth]{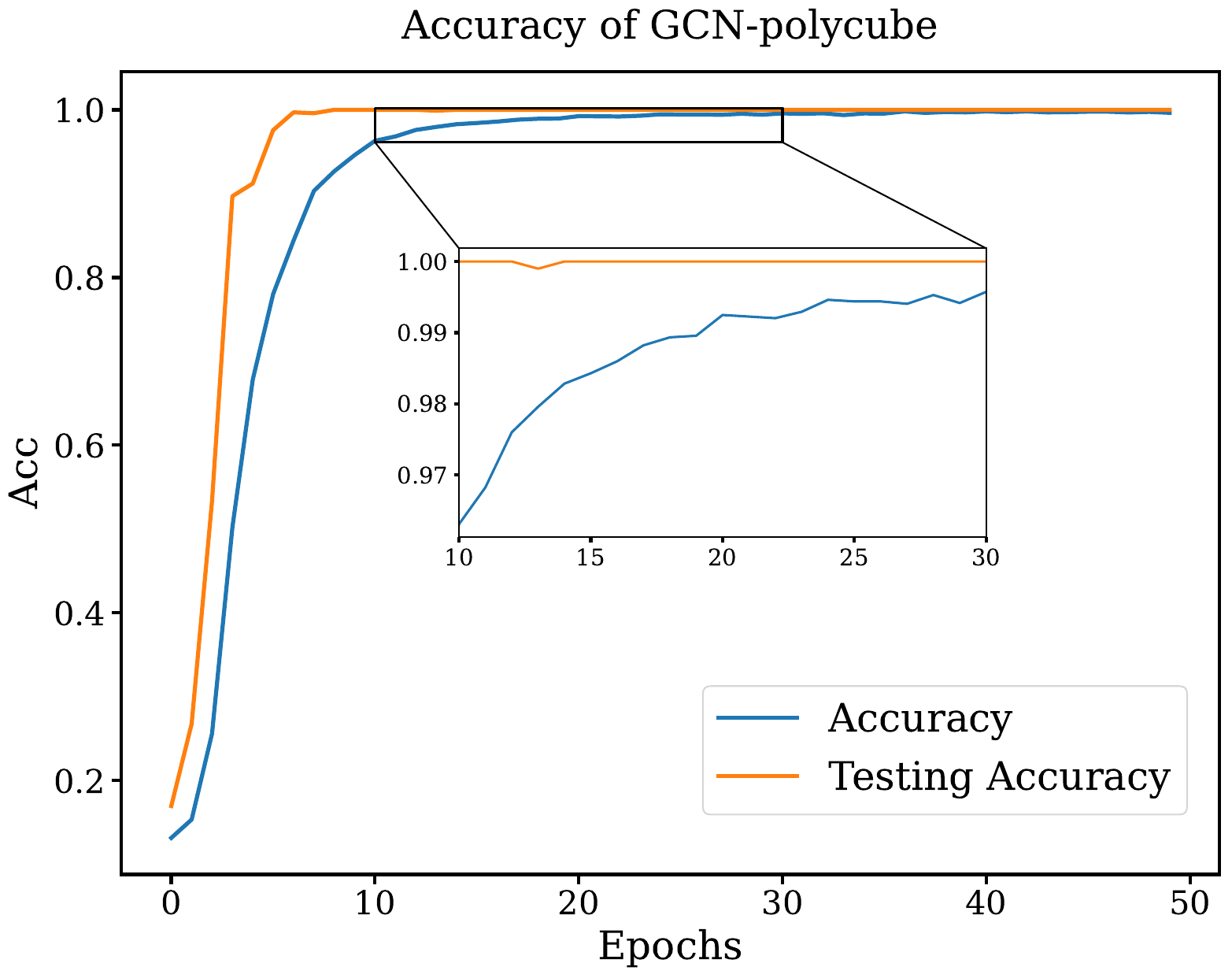}\\
  (a) & (b)\\
  \includegraphics[width=0.5\linewidth]{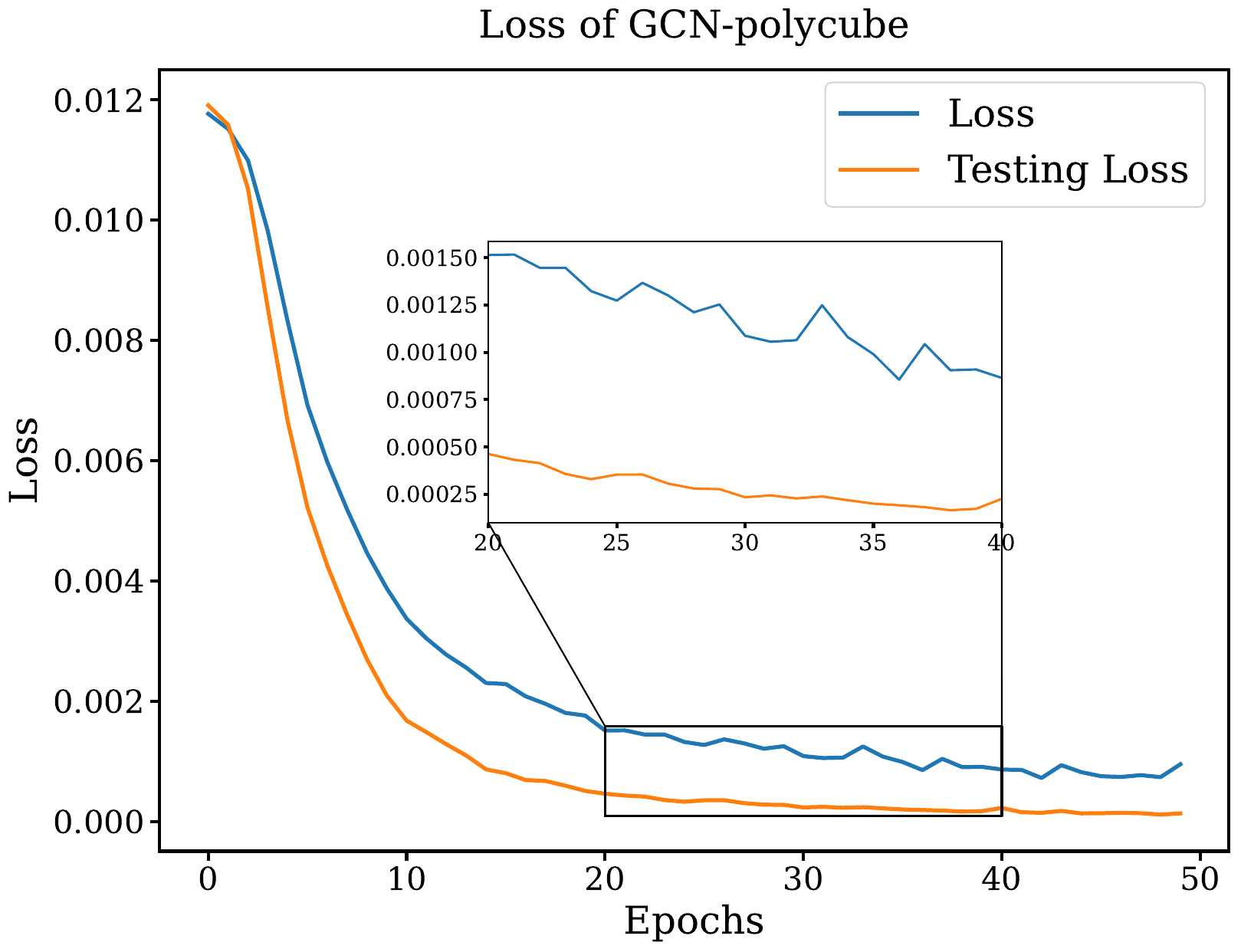}
  &\includegraphics[width=0.5\linewidth]{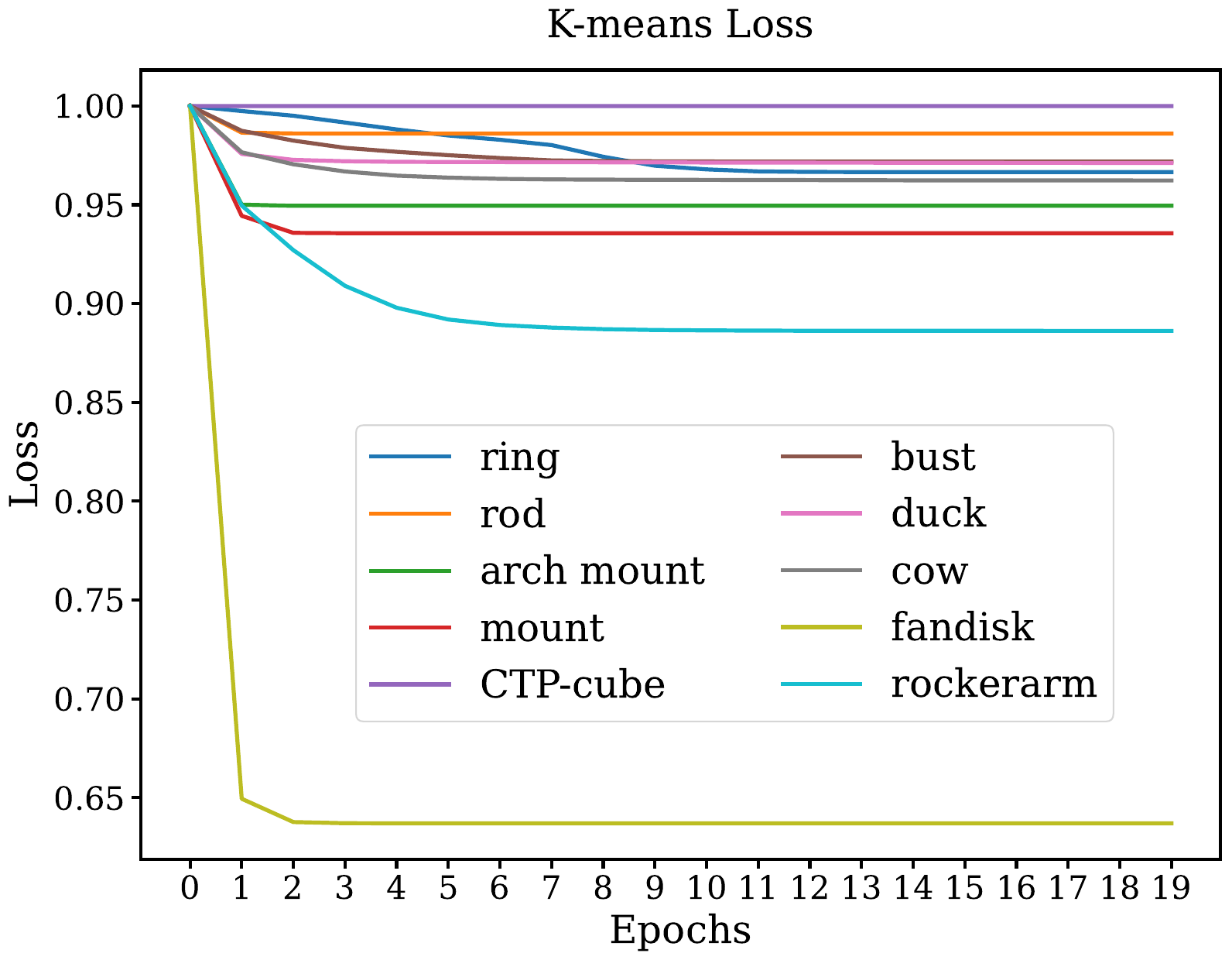}\\
  (c) & (d)\\
\end{tabular}
\caption{Hyperparameter optimization and training performance of the
  GCN-Polycube model and K-means model. (a) Hyperparameter search results for
  learning rates. (b) Accuracy of the GCN-Polycube model as a function
  of epochs. (c) Loss of the GCN-Polycube model as a function of
  epochs. (d) Loss of the K-means model. }
\label{fig:hierarch_hyperparameters}
\end{figure}

In our study, the MLP within the GCN-Polycube model consists of three hidden
layers with widths (256, 128, 128). We train the GCN-Polycube model as a deep
network using batch gradient descent and the Adam optimizer. Subsequently, we
determine the learning rate of our method through a hyperparameter search (as
shown in Fig.~\ref{fig:hierarch_hyperparameters}(a)). Through this analysis, we
identify the optimal hyperparameter setting that result in the smallest
error. Fig.~\ref{fig:hierarch_hyperparameters}(b) displays the model's accuracy
over epochs, indicating the improvement in model performance through training,
while Fig.~\ref{fig:hierarch_hyperparameters}(c) illustrates the loss function
of the GCN-Polycube model as a function of epochs, demonstrating the convergence
behavior during training.  In the training process of the K-means model, the
iteration stops when the centroids no longer change significantly (as depicted
in Fig.~\ref{fig:hierarch_hyperparameters}(d)).

\section{ML-Polycube based segmentation and path optimization}

\begin{figure}[!htb]
      \centering
      \begin{tikzpicture}
        \node[anchor=south west,inner sep=0] (image) at (0,0) {\includegraphics[width=\linewidth]{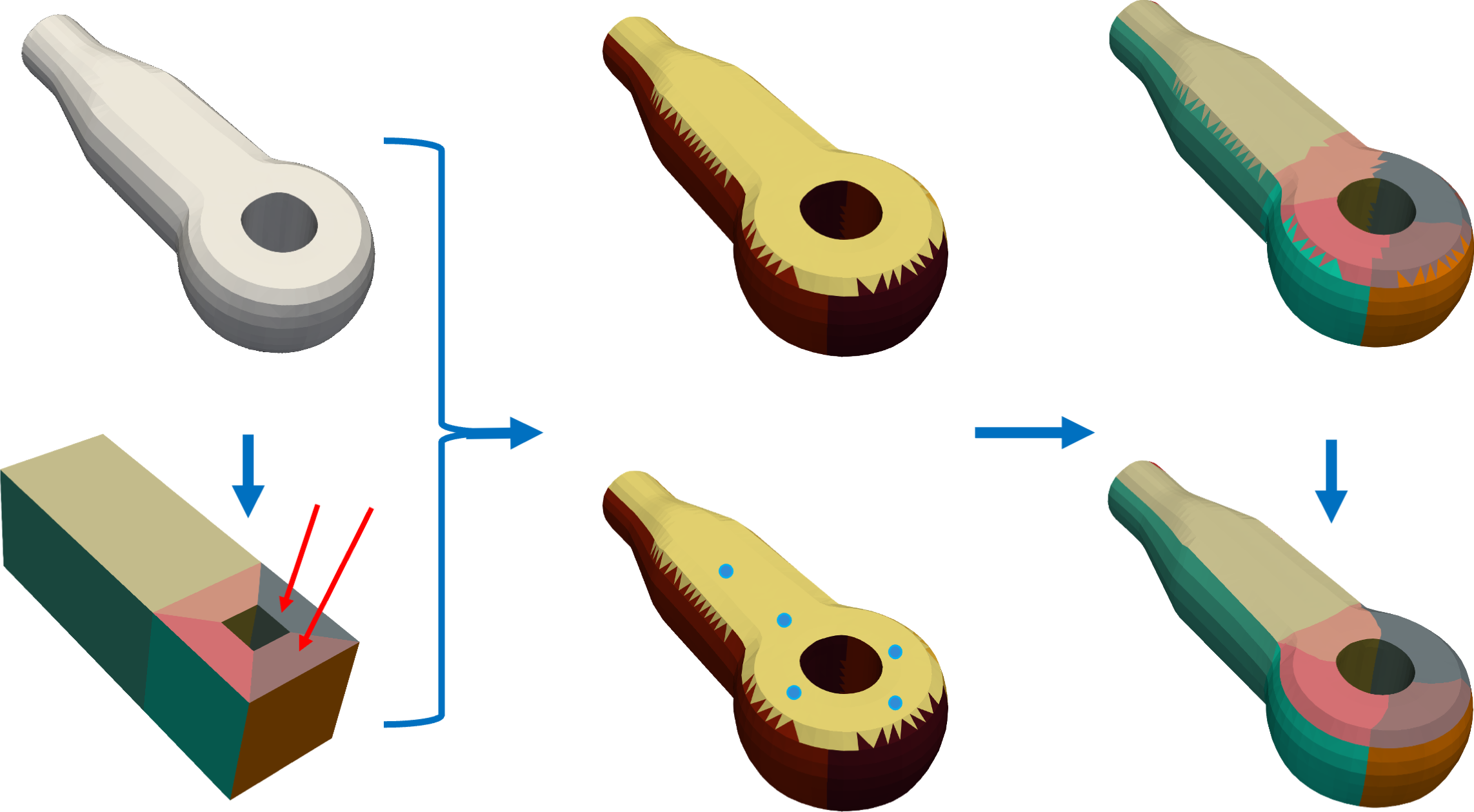}};
        \begin{scope}[x={(image.south east)},y={(image.north west)}]
          \node at (0.14,0.53) {(a)};
          \node at (0.14,-0.05) {(b)};
          \node at (0.51,0.53) {(c)};
          \node at (0.51,-0.05) {(d)};
          \node at (0.9,0.53) {(e)};
          \node at (0.9,-0.05) {(f)};
          \node at (0.22,0.43) {\scriptsize R1};
          \node at (0.26,0.43) {\scriptsize R2};
        \end{scope}
      \end{tikzpicture}
      \caption{\label{fig:Pipeline_segmentation}Surface segmentation using
        K-means clustering based on predicted polycube structure and
        GCN-centroid model. (a) The CAD geometry; (b) initial polycube structure
        prediction used to determine the number of clusters and two regions (R1
        and R2), where ``R'' stands for ``Region'', lie on the same plane within
        the polycube structure; (c) K-means clustering based on the normal space
        and the zigzag issue during segmentation; (d) feature extraction and
        centroid prediction using the GCN-centroid model; (e) K-means clustering
        in 3D Euclidean space using predicted centroids; (f)
        final optimized segmented surface.}
\end{figure}
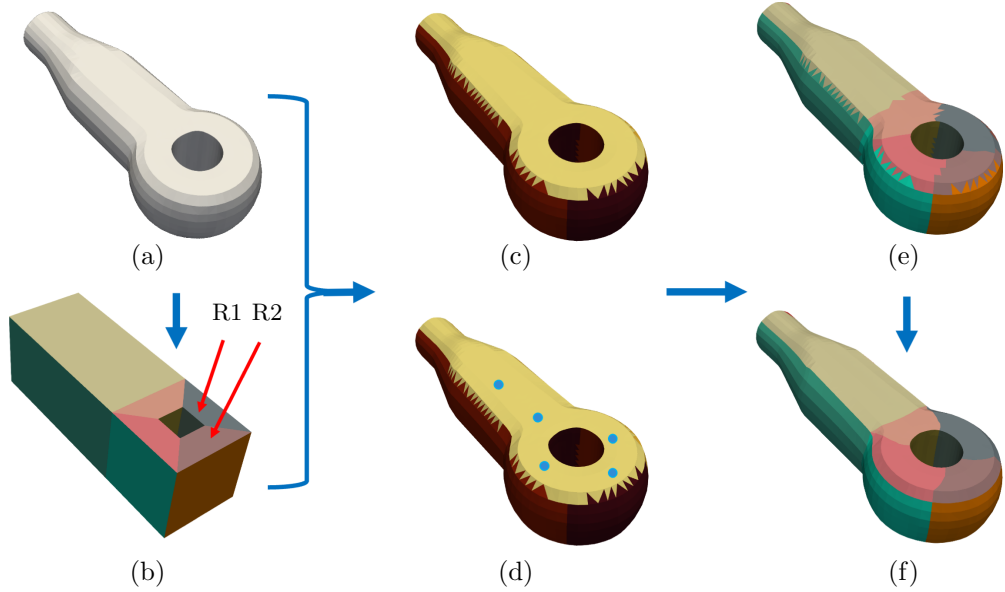

This section introduces the pipeline of integrating polycube structure
information with K-means segmentation and path optimization to produce
high-quality surface segmentation to match the polycube structure. In the
K-means step from the previous section, a significant challenge arises from the
need to segment not only based on the normal space but also based on
3D Euclidean space. It is common for two regions on the polycube
structure to lie on the same plane (see
Fig.~\ref{fig:Pipeline_segmentation}(b)). To accurately segment these regions,
especially when they are co-planar, it is important to determine the approximate
location of each cluster. This requires K-means segmentation to be performed in
both the normal space and a space formed by the centroids of the triangular
elements. A secondary GCN model (GCN-centroid) is employed to predict the
centroids when regions lie on the same plane but are separated in the polycube
structure. After segmentation based on the polycube structure is introduced, the
next challenge is the zigzag issue, as illustrated in
Fig.~\ref{fig:Pipeline_segmentation}(c). To address this problem, we use a
polycube-based Dijkstra's shortest path algorithm. The final outcome is a
one-to-one correspondence between the segmented surface and the surface of the
polycube structure.

\subsection{GCN-based centroid prediction for surface segmentation}

In the K-means step, the predicted polycube structure is used to determine the
number of clusters for the K-means algorithm. To segment the surface more
accurately, especially when two regions lie on the same plane and cannot be
distinguished using only the normal space, we use a space formed by the
centroids of the triangular elements. A GCN-centroid model is then employed to
predict these centroids. The GCN-centroid is trained to recognize and predict
the centroids based on the polycube structure, enabling the K-means algorithm to
determine the approximate location of each cluster according to the polycube
structure.

Similar to the GCN-Polycube used in polycube structure recognition, we first
perform feature extraction. Here, we calculate the centroid of a triangular
element based on its 3D position vector \(\mathbf{p}\) of each vertex
\(v_i \in \mathcal{V}\). Then we use the centroid of a triangular element as the
node feature vector. The layers of the GCN-centroid can be described by
Equation~\eqref{eq:gcn}, where \(F^{(l)}\) represents the feature matrix
corresponding to the centroid of a triangular element.

The model architecture includes four GCN-centroid layers that capture and refine
node features, which are the centroids, followed by global max pooling to
aggregate these features. Subsequent linear layers further process the pooled
features to predict the centroids. We train the GCN-centroid model using batch
gradient descent. Although both the Adam and RMSprop optimizers are suitable for
this task, we choose to use the RMSprop optimizer.

The predicted centroids are then used for K-means surface segmentation based on
3D Euclidean space. These segmentation used in conjunction
with the K-means algorithm based on the normal space to segment the surface into
\(k\) clusters, corresponding to the polycube structure. The initial centroids
for K-means are selected based on the predicted polycube structure. The K-means
segmentation process involves assigning each triangular element to the nearest
centroid based on Euclidean distance, followed by iterative refinement of the
centroids. The mathematical equation and algorithm are similar to
Equation~\eqref{eq:k-means} and Algorithm 2, but with the normal space replaced
by a space formed by the centroids of the triangular elements.

\subsection{Dijkstra’s algorithm for zigzag path optimization}

After segmenting the mesh using K-means, we often encounter zigzag issues in the
generated segments. To address this, we use Dijkstra's shortest path algorithm
to optimize the paths. The algorithm considers the geometric properties of the
edges, such as length and angles, to ensure smooth and accurate paths. By
minimizing the cumulative edge weights, the algorithm finds the shortest path in
a graph. Let $g=(v,e)$ be a graph with vertices $v$ and edges $e$, and let $w(e)$ be the
weight of edge $e$. The shortest path from vertex $v_i$ to vertex $v_j$ is
determined by:
\begin{equation}
  \text{dist}(v_i, v_j) = \min_{\text{paths } v_{ij}} \sum_{e \in v_{ij}} w(e).
\end{equation}

The path finding process involves edge weight adjustment. This adjustment
ensures that the paths found not only minimize distance but also conform to the
desired geometric constraints, such as the sharp features in the original
geometry. Here, the edge weight is adjusted as follows. For an edge \( e \)
between vertices \( v_i \) and \( v_j \), the weight \( w(e) \) is calculated
as:
\begin{equation}
  w(e) =\frac{1}{ \lambda_0} \|\mathbf{v}_i - \mathbf{v}_j\| + \lambda_1 \theta + \lambda_2 \phi,
\end{equation}
where \( \|\mathbf{v}_i - \mathbf{v}_j\| \) is the Euclidean distance. The
coefficient \( \lambda_0 \) depends on whether the edge is a sharp feature or
not. This adjustment allows the influence of sharp features to be incorporated
based on the edge length. The second term \( \theta \) considers the direction
of the current edge relative to the previous edge.  The third term \( \phi \)
controls the direction of the selected edge so that it does not deviate from the
overall direction from the origin to the destination node. The coefficients
\( \lambda_1 \) and \( \lambda_2 \) are penalty coefficients. This optimization
step ensures that the final segments align well with the polycube structure and
thus eliminates the zigzag problem.

\begin{figure}[!htb]
  \centering
  \begin{tikzpicture}
    \node[anchor=south west,inner sep=0] (image) at (0,0) {\includegraphics[width=0.8\linewidth]{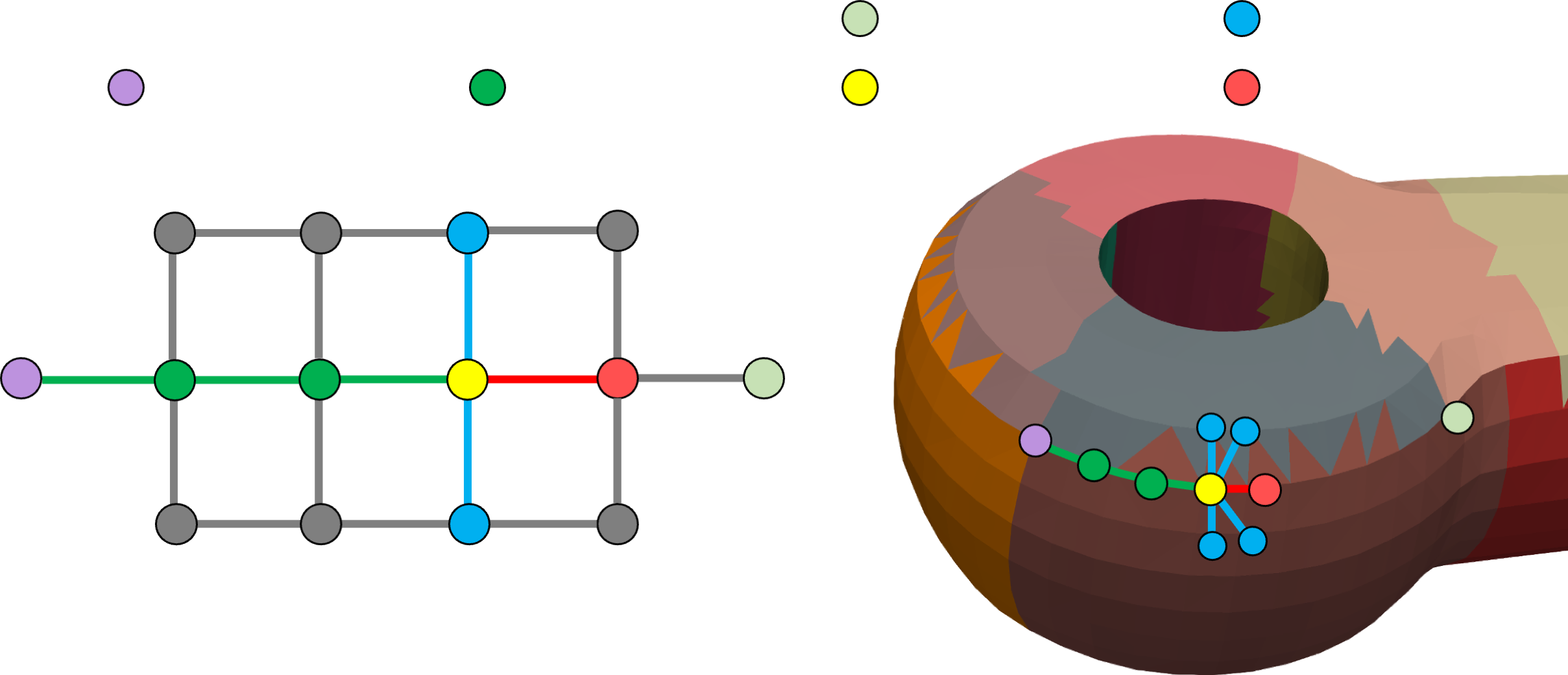}};
    \begin{scope}[x={(image.south east)},y={(image.north west)}]
      \node at (0.145,0.86) {\small Origin};
      \node at (0.425,0.87) {\small Selected node};
      \node at (0.66,0.87) {\small Current node};
      \node at (0.65,0.97) {\small Destination};
      \node at (0.88,0.87) {\small Next node};
      \node at (0.918,0.97) {\small Candidate node};
      \node at (0.24,-0.1) {(a)};
      \node at (0.75,-0.1) {(b)};
    \end{scope}
  \end{tikzpicture}
    \caption{\label{fig:hierarch_pathfinding}Pathfinding optimization using
      Dijkstra's algorithm to address zigzag issues in generated segments. (a)
      Identification of corner nodes based on the polycube structure, with a
      diagram illustrating the origin and destination nodes; (b) definition of
      potential edges between corner nodes. Calculation of edge weights
      considering Euclidean distance, angle with reference direction, and
      additional penalty terms.}
\end{figure}
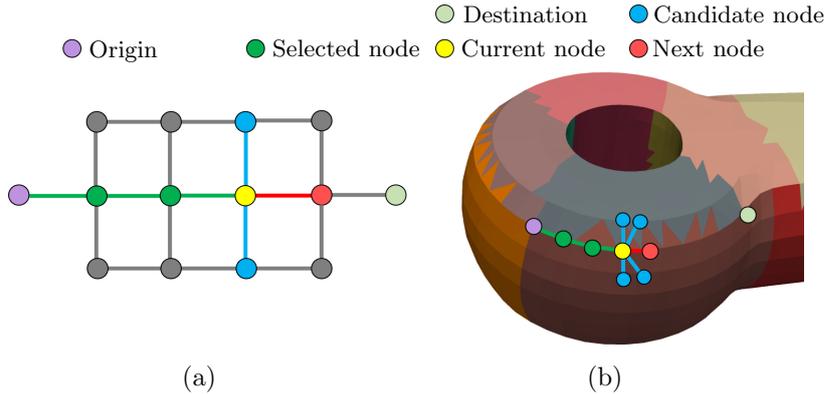

Fig.~~\ref{fig:hierarch_pathfinding} provides a detailed overview of the
algorithm used to determine the shortest path between corner nodes. The process
begins by identifying corner nodes based on polycube structure, which are then
used to define pairs of points representing potential edges. These corner pairs
are stored and used to guide the pathfinding process. The algorithm iteratively
explores paths from a starting vertex to a target vertex, updating the shortest
known path to each vertex encountered.

Once a path is found, the edges along this path are marked and stored,
preventing their reuse in subsequent iterations. This ensures that each edge is
uniquely assigned to a path, avoiding overlaps and intersections. The resulting
paths are then used to update the surface segmentation (see
Fig.~\ref{fig:Pipeline_segmentation}(e) and (f)).


\section{High-quality hex mesh generation and volumetric spline construction}
\subsection{Octree subdivision and parametric mapping}
\label{sec:octree_subdivision}

\begin{figure}[htp]
  \centering
  \begin{tabular}{cc}
    \includegraphics[height=0.45\linewidth]{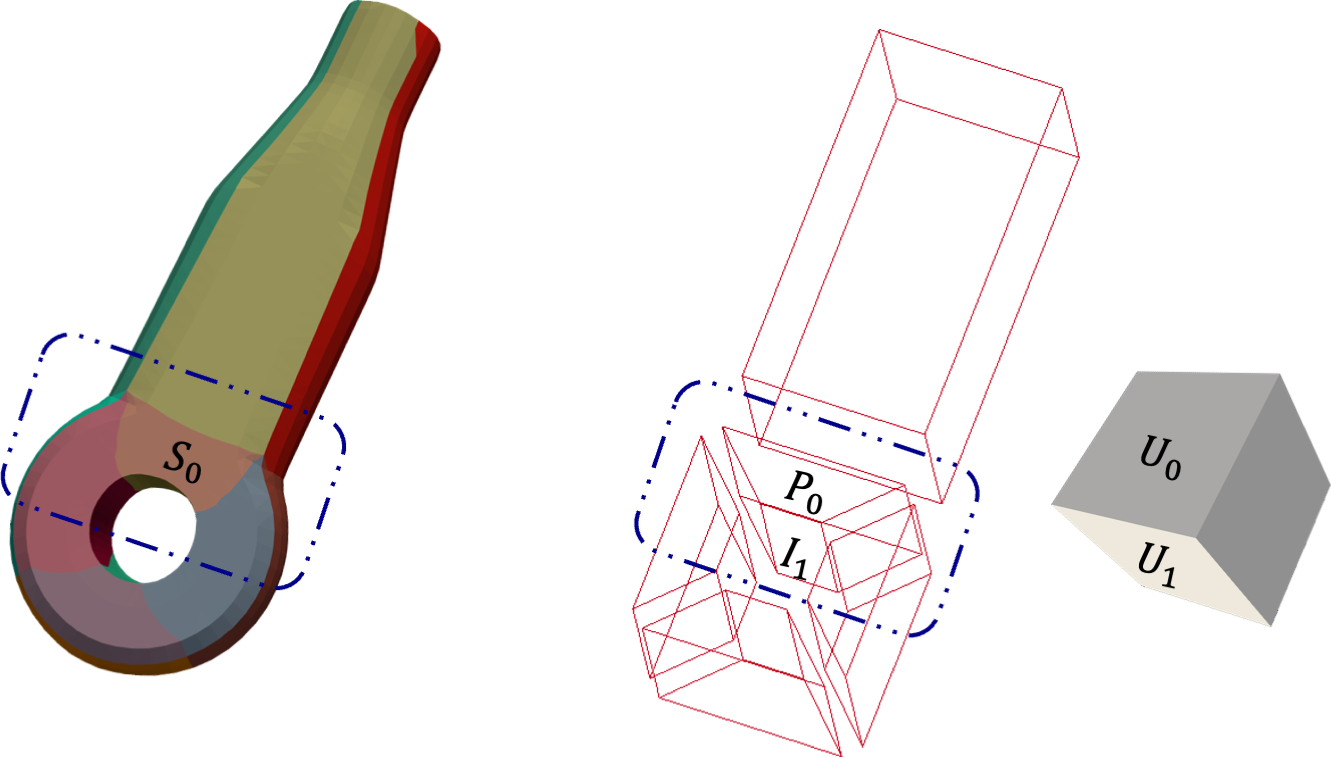}\\
    \makebox[0.2\linewidth][l]{\hspace{0.1cm}(a)}  \makebox[0.45\linewidth][c]{(b)}\\
  \end{tabular}
  \caption{The parametric mapping process for constructing an all-hex mesh using a
    predicated polycube structure. (a) The segmented surface patches
    $\{S_i\}_{i=1}^N$ derived from the K-means segmentation; (b) the polycube
    structure $\{P_i\}_{i=1}^N$ and internal surfaces $\{I_j\}_{j=1}^M$ with each
    cubic region corresponding to a unit cube $\{U_k\}_{k=1}^6$ as the parametric
    domain. The dashed rectangle highlights the correspondence between a cubic
    region in the polycube and its associated geometric volume.}
  \label{fig:rod_polycube_construction}
\end{figure}

Upon the polycube is predicted by the GCN-Polycube algorithm and the surface is
segmented by the K-means, we need to build a bijective mapping between the input
triangular mesh and the boundary surface of the polycube structure. Our
implementation adopts the strategy proposed in~\cite{Liu2015}, utilizing an
union of unit cubes as the parametric domain for the polycube structure (see
Fig.~\ref{fig:rod_polycube_construction}). The integration of the segmented
surface mesh (provided by K-means) and the polycube structure (provided by
GCN-Polycube) yield a parametric domain to perform the following octree
subdivision and parametric mapping; see the pseudocode provided in the
Parametric mapping algorithm~\cite{yu2020hexgen}.

Let $\{S_i\}_{i=1}^N$ be the segmented surface patches, derived from the
segmentation result by K-means (see
Fig.~\ref{fig:rod_polycube_construction}(a)). Each segmented surface patch
corresponds to one boundary surface of the polycube $P_i$ $(1\leq i \leq N)$, as
shown in Fig.~\ref{fig:rod_polycube_construction}(b), where $N$ represents the
number of the boundary surfaces. Additionally, the polycube includes internal
surfaces marked as $I_j$ $(1\leq j \leq M)$, with $M$ being the number of the
interior surfaces. Then, the collective set of polycube surfaces consists of
$\{P_i\}_{i=1}^N$ and $\{I_j\}_{j=1}^M$, which can be obtained automatically
from the polycube structure derived from the GCN-Polycube. For the parametric
domain, let $\{U_k\}_{k=1}^6$ denote the six surface patches of one unit cube
(see Fig.~\ref{fig:rod_polycube_construction}(b)).

Each cubic region within the polycube structure corresponds to a distinct volumetric
region of the geometry and is paired with a unit cube as its parametric domain. The
representation of a cubic region and its associated geometric volume region, highlighted
by a dashed rectangle, is shown in Fig.~\ref{fig:rod_polycube_construction}(b). Therefore,
for each cube in the polycube structure, there is a boundary surface $P_i$ to which the
segmented surface patch $S_i$ is mapped onto the corresponding parametric surface $U_k$ of
the unit cube. The mapping process from $S_i$ to $U_k$ begins by mapping the boundary
edges of $S_i$ with those of $U_k$. We then determine the parameterization of $S_i$ by
using the cotangent Laplace operator to solve for the harmonic
function~\cite{zhang_solid_2012,eck1995multiresolution}. It is important to note that the
parametric mapping step is omitted for the internal surfaces $I_j$ of the polycube.

An all-hex mesh is generated from this surface parameterization coupled with
octree subdivision. For each cubic region, vertex coordinates on the segmented
patch $S_i$ are first determined by recursively subdividing the unit cube to
acquire their parametric coordinates. The physical coordinates are then obtained
via the parametric mapping, ensuring a bijective relation between the parametric
domain $U_k$ and the physical domain $S_i$. To locate vertices on the internal
surface of the cubic section, linear interpolation is used to compute the
physical coordinates directly.  The mappings among $S_0$, $P_0$, and $U_0$ are
combined to establish the parametric mapping and to determine vertex coordinates
on the surface $S_0$. The internal surface $I_1$ is paired with $U_1$ for linear
interpolation to calculate the vertices on the internal surface of the cubic
section. Lastly, vertices within the cubic region are derived through linear
interpolation. The complete all-hex mesh is constructed by iterating over each
cubic region.

This robust component of the system, HexGen, which is part of the
HexGen\_Hex2spline subroutine, has been made publicly accessible via GitHub
(https://github.com/CMU-CBML/HexGen\_Hex2Spline.git). It comes with comprehensive
documentation and references~\cite{yu2020hexgen} for an in-depth
understanding.

\subsection{Quality metrics and quality improvement}
\label{cha:qual-meas-qual}

Mesh quality is important in both FEA and IGA, as poor quality elements can
compromise the convergence and stability of simulations.  Unfortunately, meshes
generated through the previous parametric mapping step frequently exhibit
poor-quality elements. In this section, we will present one quality metric we
take into account and three algorithms we employ for improving mesh quality.

We evaluate mesh quality based on the scaled
Jacobian~\cite{zhang2006adaptive}. For each hex element, we identify three edge
vectors $e_i = x_i - x\, (i = 0, 1, 2)$ for every corner node $x$. The Jacobian
matrix at $x$ is defined as $[e_0, e_1, e_2]$, and its Jacobian $J(x)$ is the
determinant of this matrix. We derive the scaled Jacobian $SJ(x)$ by normalizing
$e_0, e_1 \text{ and } e_2$. For each hex element, we compute the (scaled)
Jacobian at eight corners and the body center. For the body center,
$e_i\, (i = 0, 1, 2)$ are determined using three pairs of opposite face centers.

We integrate three quality improvement techniques in the software package,
namely pillowing, smoothing and optimization. Note that each quality improvement
function can be run independently and one can use these functions to improve the
mesh quality. We first use pillowing to insert one layer around the
boundary~\cite{zhang_solid_2012}. By using the pillowing technique, we ensure
that each element has at most one face on the boundary, which can help improve
the mesh quality around the boundary. To improve the mesh quality, pillowing,
smoothing, and optimization are alternately employed \cite{zhang_solid_2012,
  tong_hybridoctree_hex_2024}. This process involves moving the vertices while
ensuring that the boundary points remain on the input triangle surface. In each
iteration, optimization is carried out to the element with the worst-scale
Jacobian in the entire mesh and make adjustments to its corner
nodes. Additionally, Laplacian smoothing is applied to all vertices every $1,000$
iterations to ensure a smooth vertex distribution.

For smoothing, different relocation methods are applied to three types of
vertices: vertices on sharp edges of the boundary, vertices on the boundary
surface, and interior vertices. We perform smoothing on the surface before
smoothing the interior volume. For each sharp-edge vertex, we first detect its
two neighboring vertices on the curve, and then calculate their middle
point. For each vertex on the boundary surface, we calculate the area center of
its neighboring boundary quadrilaterals. For each interior vertex, we calculate
the weighted volume center of its neighboring hex elements as the new
position. We relocate a vertex in an iterative way. Each time the vertex moves
only a small step towards the new position and this movement is done only if the
new location results in an improved local scaled Jacobian.

In the optimization process, we observe that optimizing purely on the scaled
Jacobian function leads to gradient explosion and entrapment in local
minima~\cite{tong_hybridoctree_hex_2024}. Although optimizing purely with the
Jacobian function can solve these two problems, we cannot assess the ideal shape
of an element once the Jacobian value transitions to positive
\cite{tong_hybridoctree_hex_2024}. Therefore, we adopt a new energy function,
consisting of the geometry fitting, Jacobian, and scaled Jacobian terms. We have
\begin{equation}
\label{equ:1}
E = \sum_{i=0}^{\mathit{nn_s} - 1} \lVert x_i - x_i^s \rVert_2^2  - \frac{1}{\bar{l}}\sum_{{j} = 0}^{{\mathit{ne_n}} - 1} \min\mathit{J}(h_{{j}}) - \bar{l}^2 \sum_{{k} = 0}^{{\mathit{ne_p}} - 1} \min SJ(h_{{k}}),
\end{equation}
where $\mathit{nn_s}$ is the number of surface vertices, $\mathit{ne_n}$ is the
number of hex elements with negative Jacobians, $\mathit{ne_p}$ is the number of
hex elements with positive Jacobians, and $\bar{l}$ is the average length of
three edges to calculate the (scaled) Jacobian. Let us assume the fundamental
dimension of geometric length is \(L\). Then the dimension of the geometry
fitting term is \(L^2\), the Jacobian and the scaled Jacobian are in the
dimension of $L^3$ and $L^0$, respectively. To unify the dimension of the
Jacobian and scaled Jacobian terms to \(L^2\), we introduce $\frac{1}{\bar{l}}$
to the Jacobian term and $\bar{l}^2$ to the scaled Jacobian term. We adopt the
gradient-based method to iterative minimize the energy function. All the mesh
vertices are optimized by
\begin{equation}
x_i \rightarrow x_i - \alpha\nabla E_{x_i}, \quad i = 0, 1, \dots, \mathit{nn} - 1,
\end{equation}
where $\mathit{nn}$ is the number of all vertices. We choose the weight
$\alpha = 1 \times 10^{-4}$ for all the tested models in this paper.

To efficiently determine the nearest surface point $x_i^s$ for each boundary
point $x_i$, we check all triangles within a bounding box ten times the length
of the local triangular edge. The identification process for $x_i^s$ varies
depending on whether $x_i$ is projected onto a face, edge, or point. If $x_i$ is
projected onto a face, we iterate through all triangles in the bounding box,
calculate the point-to-triangle distance, and determine the nearest projection
point $x_i^s$. In contrast, if $x_i$ is directly projected onto a point, $x_i^s$
simply becomes that point. If $x_i$ is projected onto an edge, $x_i^s$ is
searched on the corresponding sharp edges in the triangle mesh. To expedite the
process, we update the closest triangle index for each $x_i$ every 1,000
iterations. If the maximum of the minimum distances from all points to the
surface is less than $10^{-8}$ of the overall bounding box edge size, we
directly pull each $x_i$ to its corresponding $x^s$. We present Algorithm 3 for
the entire quality improvement pipeline.
\begin{algorithm}[H]
  \caption{\textbf{Quality Improvement Algorithm} }
  \label{alg:4}
        \begin{algorithmic}[1]
\State \textbf{Input:} Manifold, watertight triangular mesh $T$ with annotated sharp features, an all-hex mesh $\mathcal{H}$ to be fitted to $T$
\State Initialize $N \gets \textit{\#elem}\in\mathcal{H}, \alpha\gets 10^{-4}$
\For{$x_i\in\mathcal{H}$ surface}
    \If{$x_i$ is closest to a corner node}
        \State Classify $x_i$ as a corner node
    \ElsIf{A* algorithm determines $x_i$ is along a shortest path}
        \State Classify $x_i$ as an edge point
    \Else
        \State Classify $x_i$ as a face point
    \EndIf
\EndFor
\While {minimum scaled Jacobian $<$ given threshold}
    \State Calculate $x_i^s, \forall x_i\in\mathcal{H}$ surface
    \State // Gradient-Based Mesh Quality Optimization
    \State $g \leftarrow \nabla E_{x_i}$
    \State $x_i \gets x_i + \alpha g$\Comment{Update vertex position based on Jacobian}
    \State // Laplacian Smoothing
    \If {iter$\ \%\ 1000 == 0$}
        \State smartLaplacianSmoothing($\mathcal{H}$)\Comment{Smooth vertex position}
    \EndIf
    \State iter$\ \gets\ $iter$\ +\ 1$
\EndWhile
\State \textbf{return} $\mathcal{H}$
\end{algorithmic}
\end{algorithm}

\subsection{{Hex2Spline: volumetric spline construction}}
\begin{figure}[!htb]
  \centering
  \begin{tikzpicture}
    \node[anchor=south west,inner sep=0] (image) at (0,0) {\includegraphics[width=0.8\linewidth]{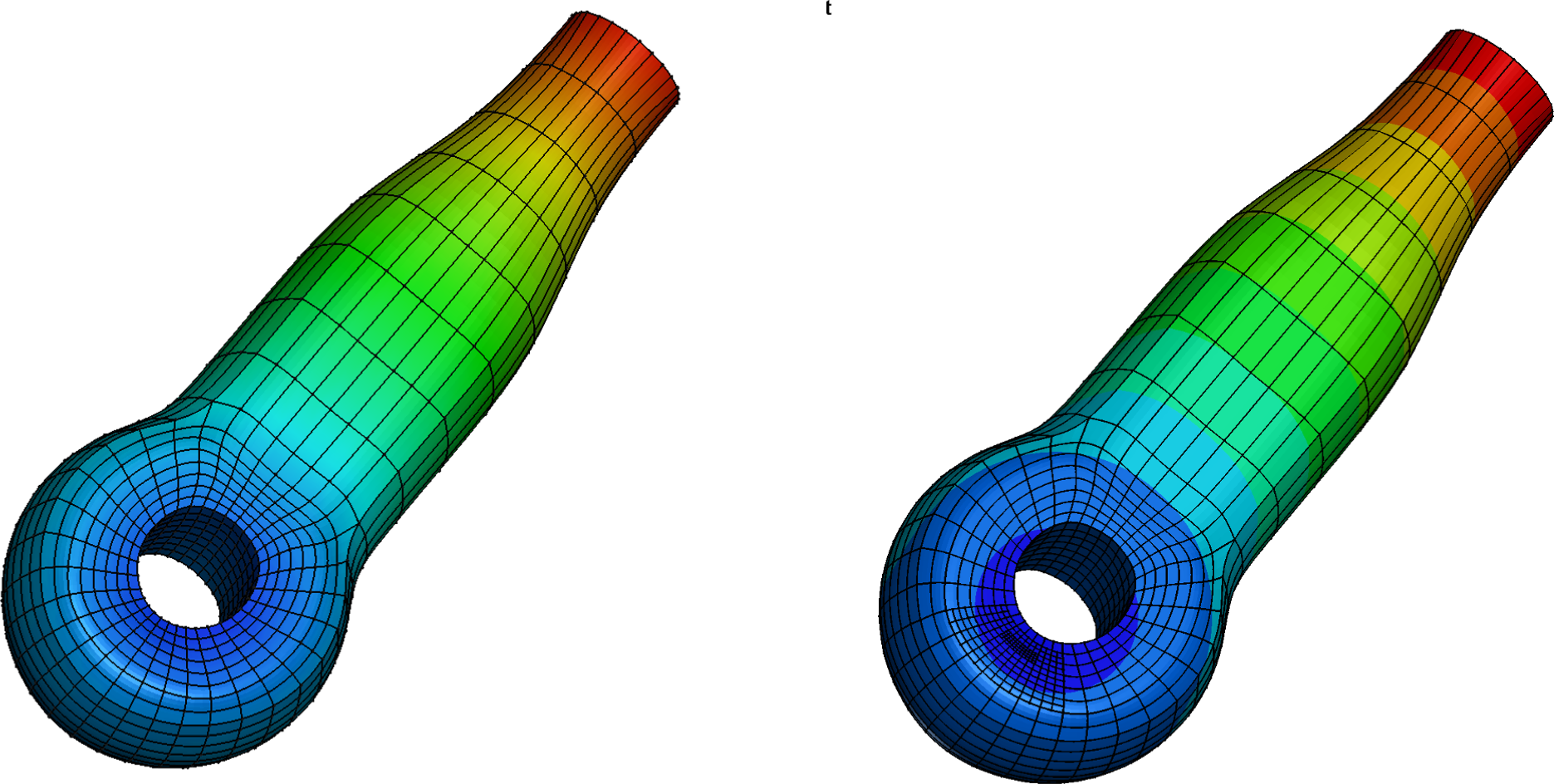}};
    \begin{scope}[x={(image.south east)},y={(image.north west)}]
      \node at (0.15,-0.05) {(a)};
      \node at (0.75,-0.05) {(b)};
    \end{scope}
  \end{tikzpicture}
  \caption{\label{fig:IGA_sample} Visualization of the output B\'{e}zier mesh
    with: (a) global refinement and (b) two levels of local refinement with IGA
    eigenvalue analysis in ANSYS-DYNA.}
\end{figure}

Upon acquiring the hex mesh of satisfactory quality, the subsequent step
involves constructing the volumetric spline, specifically the TH-spline3D, on
these unstructured hex meshes. This process consists of two steps: the
construction of TH-spline3D on hex meshes and the extraction of B\'{e}zier
elements for IGA simulation in ANSYS-DYNA. Our previous software, Hex2Spline,
uses a generated hex mesh as the input control mesh to facilitate the
construction of a TH-spline3D over CAD geometry. This robust portion of the
system has been made publicly accessible via GitHub
(https://github.com/CMU-CBML/HexGen\_Hex2Spline.git), complete with
comprehensive documentation and reference~\cite{yu2020hexgen,wei17a} for an
in-depth understanding.  Capable of defining spline functions over arbitrarily
unstructured hex meshes, Hex2spline supports sharp feature preservation, and
global refinement (see Fig.~\ref{fig:IGA_sample}(a)), Hex2spline also
incorporates adaptive IGA with its local refinement capabilities (see
Fig.~\ref{fig:IGA_sample}(b). Hex2Spline can output the B\'{e}zier information
of constructed volumetric splines for IGA simulation in ANSYS-DYNA or
alternative IGA platforms.

\section{Results}
\label{sec:exper-proc-results}
In this section, we demonstrate the effectiveness and robustness of the
DL-Polycube algorithm through various test cases. We evaluate the performance of
our approach on a range of CAD geometries with varying complexities and
topologies. The results are analyzed in terms of polycube structure prediction
accuracy, and the overall quality of the generated hex meshes and
IGA results.

\subsection{Dataset of deep learning models}
\label{sec:dataset}

\begin{figure}[!htb]
  \center{\includegraphics[width=\linewidth]{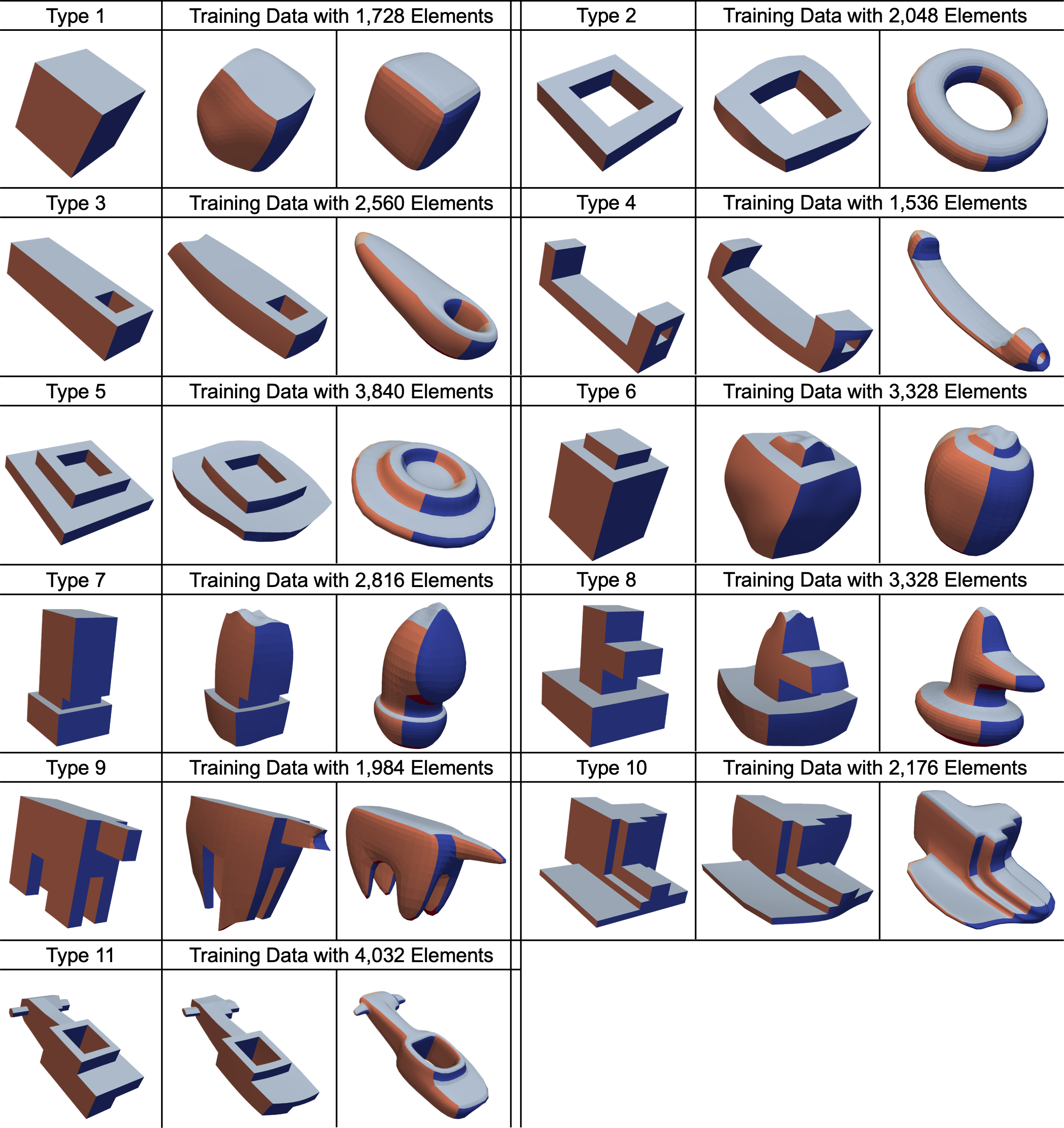}}
  \caption{\label{fig:training_data} Visualization of the training dataset for
    the deep learning model. Polycube structures are shown on the left for each type, ordered by
    complexity, along with two randomly selected derivative geometries for each
    polycube type.}
\end{figure}

The training process for the deep learning models is conducted using a large
dataset of generated geometric models. The dataset includes a wide variety of
polycube structures and their corresponding deformed geometries, ensuring
diverse training examples. The dataset was generated using Blender and its
built-in Python and BMesh libraries. We created 11 types of polycube structures,
each with 900 derivative geometries, resulting in a total of 9,900 training
examples. Fig.~\ref{fig:training_data} displays the dataset, including the
polycube structure and their derivative geometries, with two randomly chosen
examples.

\subsection{Polycube structure prediction}
\label{sec:polyc-struct-pred}

\begin{figure}[!htp]
  \centering
  \begin{tabular}{c}
    \includegraphics[width=\linewidth]{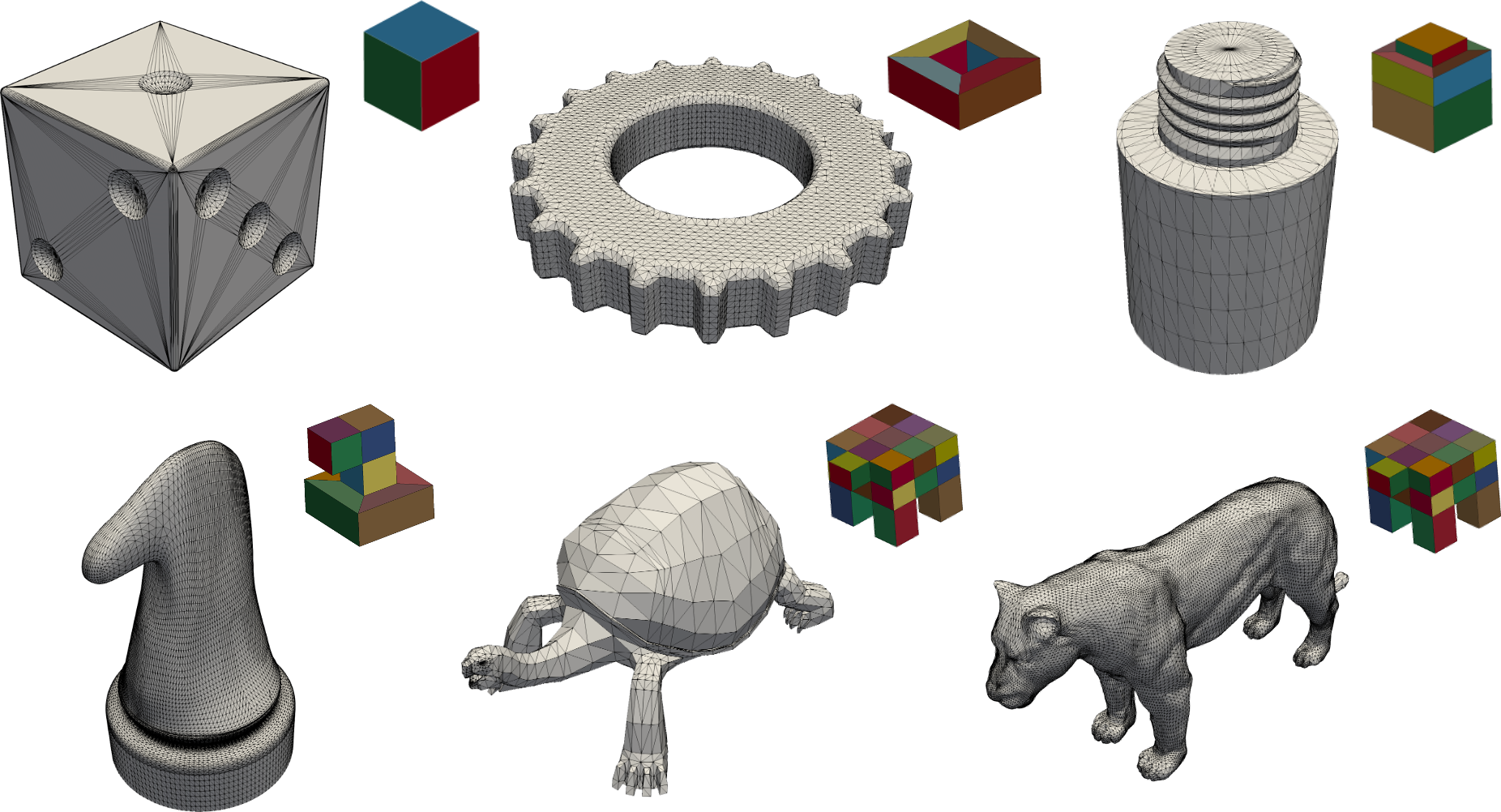}
  \end{tabular}
  \caption{ Surface triangle meshes with predicted polycube structures for the additional six models. }
  \label{fig:model4}
\end{figure}

We tested the DL-Polycube algorithm on a set of 16 CAD models, including
mechanical parts and free-form geometries. We applied deep learning to classify
the polycube structures of these 16 models in order to determine the most likely
polycube structure. The accuracy of polycube structure prediction was evaluated
by comparing the predicted structures with the constructed ground truth polycube
structures. The probability distributions output by the model are shown in
Table~\ref{Polycube_Table_1}. To clarify, we have named Type-1 as the P1 column,
Type-2 as the P2 column, Type-3 as the P3 column, and so on. The order of types
corresponds to Fig.~\ref{fig:training_data}. The probability distribution
represents the model's confidence in how well the predicted polycube structure
matches each type of the ground truth. Specifically, for the rod model, the
probability for P3 (Type-3) is 100.00\%, which indicates that the model believes
the predicted structure most likely belongs to Type-3. The surface triangle
meshes with predicted polycube structures for these models are shown in
Figs.~\ref{fig:model1},~\ref{fig:model2},~\ref{fig:model3},
and~\ref{fig:model4}. Interestingly, the probability distribution of predicted
polycube structures shows that some models can potentially match multiple
polycube structures. In such cases, we select the polycube structure
corresponding to the highest probability. For example, in the case of the dice
model, it has a 65.13\% probability for P1 (Type-1), but also a 34.6\%
probability for P6 (Type-6). Here, we choose the structure with the highest
probability, which is Type-1.

\begin{table}[htp]
  \caption{Probability distribution of predicted polycube structures for 16 CAD models.}
\label{Polycube_Table_1}
\centering
\setlength{\tabcolsep}{1.7pt}
\scriptsize
\begin{tabular}{c|c|c|c|c|c|c|c|c|c|c|c}
  \hline
  Model &P1 & P2 & P3 & P4 & P5 & P6 & P7 & P8 & P9 & P10 &P11  \\
  \hline
  ring (Fig.~\ref{fig:Pipeline_Structure})&0.00 &\textbf{99.92} & 0.00 & 0.00 & 0.08 & 0.00 & 0.00 & 0.00 & 0.00 & 0.00 & 0.00 \\
  rod (Fig.~\ref{fig:model1})&0.00&0.00& \textbf{100.00} & 0.00 & 0.00 & 0.00 & 0.00 & 0.00 & 0.00 & 0.00 & 0.00 \\
  arch mount (Fig.~\ref{fig:model1})&0.00 & 0.00 & 0.01 & \textbf{95.92} & 0.00 & 0.00 & 0.00 & 0.00 & 0.14 & 0.00 & 3.93  \\
  mount  (Fig.~\ref{fig:model1})&0.00 & 0.12 & 0.04 & 0.00 &\textbf{99.82} & 0.01 & 0.00 & 0.00 & 0.00 & 0.00 & 0.01  \\
  CTP-cube (Fig.~\ref{fig:model2})&0.79 & 0.00 & 1.61 & 0.00 & 0.06 &\textbf{91.79} & 2.99 & 2.56 & 0.20 & 0.00 & 0.00  \\
  bust (Fig.~\ref{fig:model2})&0.14 & 0.00 & 0.00 & 0.00 & 0.00 & 5.26 & \textbf{89.60} & 0.00 & 5.00 & 0.00 & 0.00 \\
  duck (Fig.~\ref{fig:model2})&0.00 & 0.00 & 14.91 & 0.00 & 0.00 & 1.93 & 0.01 & \textbf{83.15} & 0.00 & 0.00 & 0.00 \\
  cow (Fig.~\ref{fig:model3})&0.00 & 0.00 & 0.00 & 0.00 & 0.00 & 0.00 & 0.00 & 0.00 & \textbf{100.00} & 0.00 & 0.00 \\
  fandisk (Fig.~\ref{fig:model3})&0.00 & 0.00 & 0.00 & 0.00 & 0.00 & 0.02 & 0.00 & 0.00 & 0.03 &\textbf{99.75} & 0.20 \\
  rockerarm (Fig.~\ref{fig:model3})&0.00 & 0.00 & 0.05 & 0.01 & 0.71 & 0.02 &0.00 & 0.00 & 0.03 & 0.00 &\textbf{99.18}  \\
  dice (Fig.~\ref{fig:model4})&\textbf{65.13} &0.00 & 0.00 & 0.00 & 0.00 & 34.60 & 0.27 & 0.00 & 0.00 & 0.00 & 0.00 \\
  cog (Fig.~\ref{fig:model4})&0.00&\textbf{97.82} & 0.00 & 0.00 & 2.10 & 0.00 & 0.00 & 0.00 & 0.00 & 0.00 & 0.08 \\
  screw (Fig.~\ref{fig:model4})&0.00 & 0.00 & 0.03 & 0.00 & 0.00 & \textbf{99.75} & 0.02 & 0.20 & 0.00 & 0.00 & 0.00  \\
  knight  (Fig.~\ref{fig:model4})&0.26 & 0.00 & 0.00 & 0.00 &0.00 & 0.13 & \textbf{99.55} & 0.00 & 0.06 & 0.00 & 0.00  \\
  turtle (Fig.~\ref{fig:model4})&0.00 & 0.00 & 7.69 & 0.05 & 0.01 &0.13 & 0.01 & 0.01 & \textbf{91.61} & 0.00 & 0.49  \\
  lion (Fig.~\ref{fig:model4})&0.00 & 0.00 & 2.56 & 0.00 & 0.00 &0.08 & 0.00 & 0.00 & \textbf{97.36} & 0.00 & 0.00  \\
  \hline
\end{tabular}
\end{table}

\subsection{Hex mesh generation and volumetric spline construction}
\label{sec:hexah-mesh-gener}

The Polycube structure prediction and surface segmentation are performed using
the DL-Polycube program.  The DL-Polycube algorithms are implemented in Python
using the PyG library~\cite{fey2019fast}. The software is open-source and available at the
following GitHub repository: https://github.com/CMU-CBML/DL-polycube. The hex
mesh generation and volumetric spline construction are done with the help of two
software packages: Hex2Gen and Hex2Spline. These packages are open-source and
available at the following GitHub repository:
https://github.com/CMU-CBML/HexGen\_Hex2Spline.  The hex mesh generation and
volumetric spline construction algorithms are implemented in C++, utilizing the
Eigen library~\cite{eigenweb} and Intel MKL~\cite{intel-alt} for matrix and
vector operations and numerical linear algebra.

\begin{landscape}
\begin{figure}[!htp]
\centering
\begin{tabular}{l}
  \includegraphics[width=\linewidth]{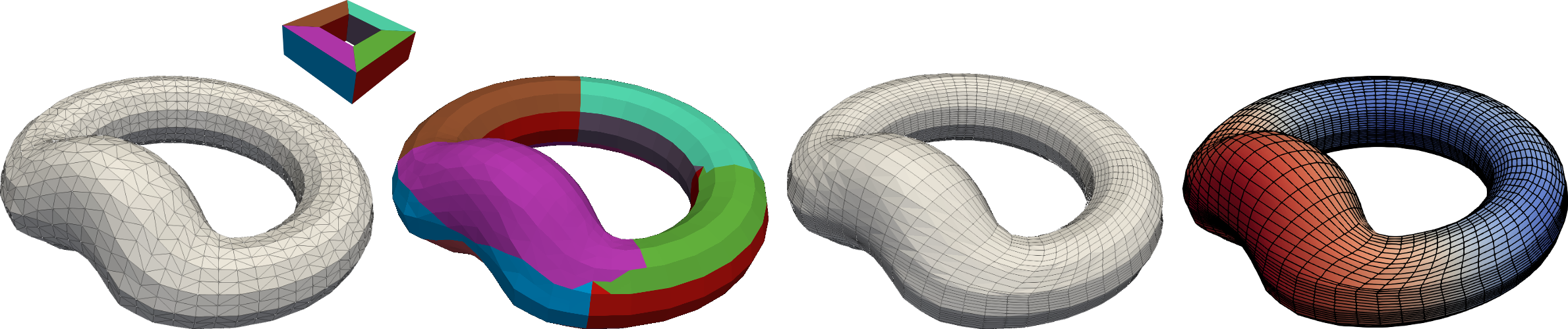}\\
  \includegraphics[width=\linewidth]{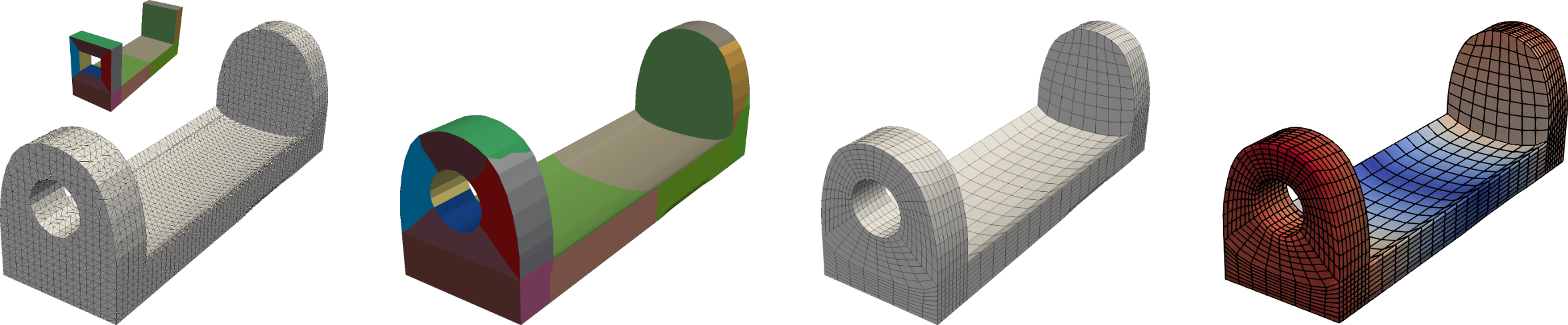}\\
  \begin{tikzpicture}
    \node[anchor=south west,inner sep=0] (image) at (0,0) {\includegraphics[width=\linewidth]{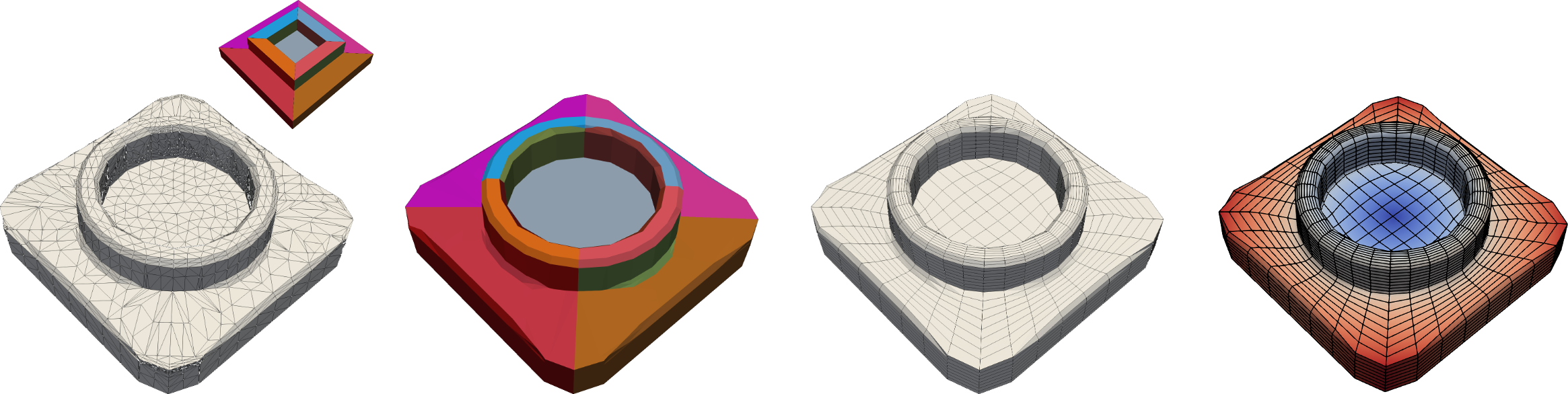}};
    \begin{scope}[x={(image.south east)},y={(image.north west)}]
      \node at (0.11,-0.1) {(a)};
      \node at (0.37,-0.1) {(b)};
      \node at (0.63,-0.1) {(c)};
      \node at (0.89,-0.1) {(d)};
    \end{scope}
  \end{tikzpicture}
\end{tabular}
\caption{ Results of ring, arch mount, and mount models. (a) Surface triangle meshes
  with its predicted polycube structures; (b) segmentation results; (c) all-hex
  control meshes; (d) volumetric splines with IGA eigenvalue analysis
  in ANSYS-DYNA. }
    \label{fig:model1}
\end{figure}
\end{landscape}

We have applied GCN, K-means, and two software packages to various models in a
fully automated process. The results were computed on a PC equipped with a 3.1
GHz Intel Xeon w5-2445 CPU, 64 GB of RAM, a 16 GB GPU, and 32 GB of shared
memory with an RTX 4080. The polycube structure is predicted using GCN-Polycube,
while surface segmentation is based on K-means. If necessary, GCN-centroid is
used to resolve segmentation issues that cannot be matched to the polycube
structure when using only normal space in K-means. To address the zigzag
problem, we employed Dijkstra’s algorithm. Some initial triangular meshes have
connectivity issues: if certain parts of the mesh are not connected, the
Dijkstra algorithm may fail to find a smooth path from start to
finish. Therefore, after identifying the polycube structure and encountering
zigzag issues during segmentation, we refine the initial triangular mesh to
resolve these connectivity problems. The predicted polycube structure and
surface segmentation are then used as inputs for HexGen and Hex2Spline, which
generate high-quality all-hex meshes and splines.  For selected ten models, we
present the predicated polycube, surface segmentation, and the final all-hex
mesh. The models include rod model (Fig.~\ref{fig:Pipeline_Structure}); ring,
arch mount, and mount models (Fig.~\ref{fig:model1}); cylinder-topped perforated
cube (CTP-cube), bust, and duck models (Fig.~\ref{fig:model2}); and cow,
fandisk, and rockerarm models (Fig.~\ref{fig:model3}). Table~\ref{hex_Table_3}
provides statistics of all tested models. The quality of the all-hex meshes is
assessed using the scaled Jacobian, the obtained meshes exhibit good quality
(minimal Jacobian $> 0.1$).

After generating all-hex meshes, we tested these ten models for IGA using
TH-spline3D. The result of the spline construction is $C^0$-continuous around
extraordinary points and edges, while maintaining $C^2$-continuous in all other
regions. Subsequently, B\'{e}zier elements are extracted for the IGA
analysis. For each testing model, we use ANSYS-DYNA to perform eigenvalue
analysis and show the first mode result
(Figs.~\ref{fig:Pipeline_Structure},~\ref{fig:model1},~\ref{fig:model2},
and~\ref{fig:model3}). The results indicate that our algorithm successfully
produces valid volumetric spline structures for IGA applications in ANSYS-DYNA.

\begin{table}[htp]
\caption{Statistics of the tested models for hex mesh generation.}
\label{hex_Table_3}
\centering
\setlength{\tabcolsep}{3.7pt}
\scriptsize
\begin{tabular}{c|c|ccc}
  \hline
  Model &Input triangle mesh &Octree &Output hex mesh&Worst\\
        &(vertices elements)&levels&(vertices elements)&Jacobian\\
  \hline
  ring (Fig.~\ref{fig:model1})           &(1,536 3,072)    &4 &(22,592 20,480)&0.54\\
  rod (Fig.~\ref{fig:Pipeline_Structure})&(11,436 22,872)  &3 &(4,520 3,840) &0.28 \\
  arch mount (Fig.~\ref{fig:model1})     &(38,592 77,184)  &3 &(5,832 4,608) &0.19 \\
  mount (Fig.~\ref{fig:model1})          &(13,136 26,268)  &3 &(7,641 6,656) &0.16 \\
  CTP-cube (Fig.~\ref{fig:model2})       &(1,122 2,240)    &3 &(9,317 8,448) &0.44 \\
  bust (Fig.~\ref{fig:model2})           &(8,540 17,076)   &3 &(42,851 39,424) &0.20 \\
  duck (Fig.~\ref{fig:model2})           &(12,217 24,372)  &3 &(7,547 6,528) &0.29 \\
  cow  (Fig.~\ref{fig:model3})           &(3,359 6,714)    &3 &(29,745 27,648) &0.61 \\
  fandisk (Fig.~\ref{fig:model3})        &(25,894 51,784)  &3 &(33,001 30,720) &0.19 \\
  rockerarm (Fig.~\ref{fig:model3})      &(69,936 139,872) &3 &(68,136 64,000) &0.12 \\
  \hline
\end{tabular}
\end{table}

\section{Conclusions and future work}

In this paper, we present a novel approach called DL-Polycube, which integrates
deep learning with the polycube method to automate the generation of
high-quality hex meshes and volumetric splines. By incorporating deep learning,
we significantly reduce the manual effort required for surface segmentation and
polycube construction in our previously developed software packages
\cite{yu2020hexgen} and improve correction procedures to address the issue where
not every labeling permits a corresponding polycube, as mentioned in
\cite{pietroni_hex-mesh_2023}. Since the labeling is automatically generated
from the predicted polycube structure, these labelings naturally correspond to a
polycube. Our approach leverages machine learning to automate the construction
of polycube structures from CAD geometries and the subsequent surface
segmentation. The surfaces of the predicted polycube structures and the
segmented CAD geometry maintain a one-to-one correspondence. This algorithm not
only automates the conversion of CAD geometries into high-quality hex meshes but
also reduces the learning curve for users, allowing them to quickly become
familiar with in generating hex meshes and constructing volumetric splines. The
robustness and efficiency of the DL-Polycube algorithm are demonstrated through
several examples.

While the DL-Polycube algorithm represents a significant advancement in our
early work on hex mesh generation and volumetric spline construction, there are
several areas for future improvement. First, it is important to note that the
variability of the design parameters and topologies within the dataset
constrains the generalizability of the machine learning model. If the geometry
model encounters design parameters and topologies outside its trained range, it
may not predict an ideal polycube structure, leading to poor-quality hex
mesh. Therefore, we plan to include a wider variety of design parameters and
topologies in our dataset to cover more geometric configurations. A geometric
library based on more polycubes can be established to facilitate this. Second,
this paper focuses on genus-0 and genus-1 geometries. For high-genus geometries
and complex genus-0 and genus-1 geometries, generative models can be employed to
automatically separate them into multiple simpler genus-0 and genus-1
geometries, each simpler genus-0 or genus-1 geometry can then be processed using
the methods described in this paper. This approach will enable the generation of
hex meshes and the construction of splines for high-genus and complex
geometries, which is an area of interest for us.  Third, although our focus has
been on hex meshes, extending the DL-Polycube algorithm to generate other types
of meshes, such as hex-dominant meshes, could broaden its applicability.

In conclusion, the DL-Polycube algorithm represents a significant step forward
in automating the hex mesh generation and volumetric spline construction. By
addressing the future work, we aim to further enhance the capabilities and
applicability of our approach, ultimately contributing to the advancement of hex
mesh generation and IGA.

\bmhead{Acknowledgements}

This project is supported by the National Natural Science Foundation of China
(Grant No. 62302091), the Discipline Innovation Field Cultivation Project (Grant
No. XKCX202308), and the Fundamental Research Funds for the Central Universities
(Grant No. 2232023D-24). H. Tong and Y. J. Zhang were supported in part by a
Honda project.

\begin{landscape}
\begin{figure}[!htp]
\centering
\begin{tabular}{l}
  \includegraphics[width=0.8\linewidth]{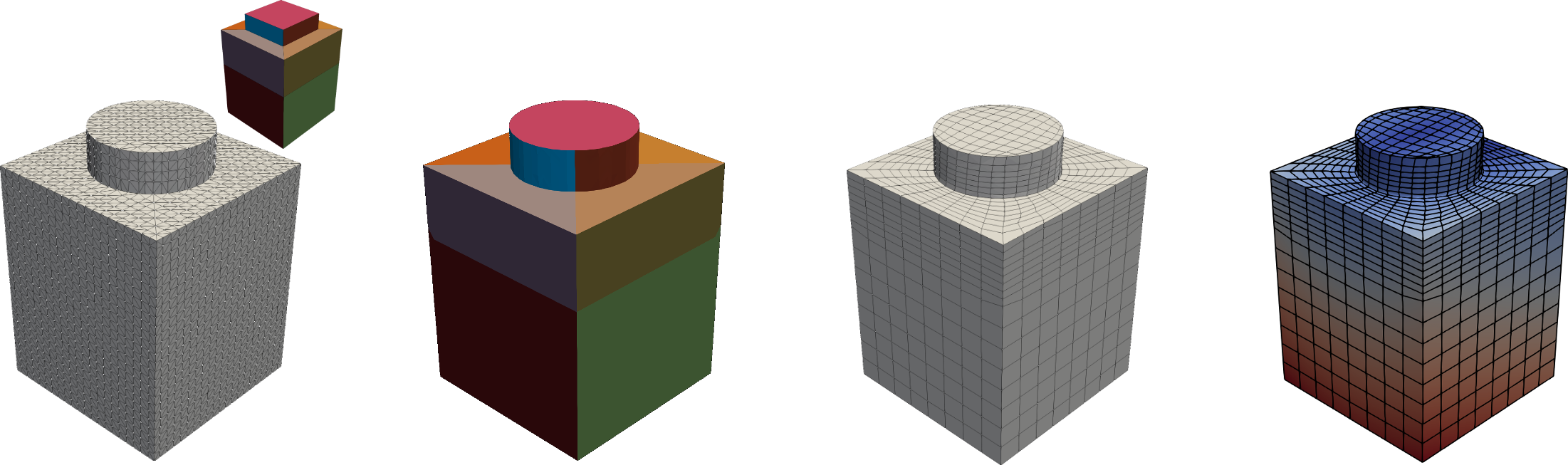}\\
  \includegraphics[width=0.8\linewidth]{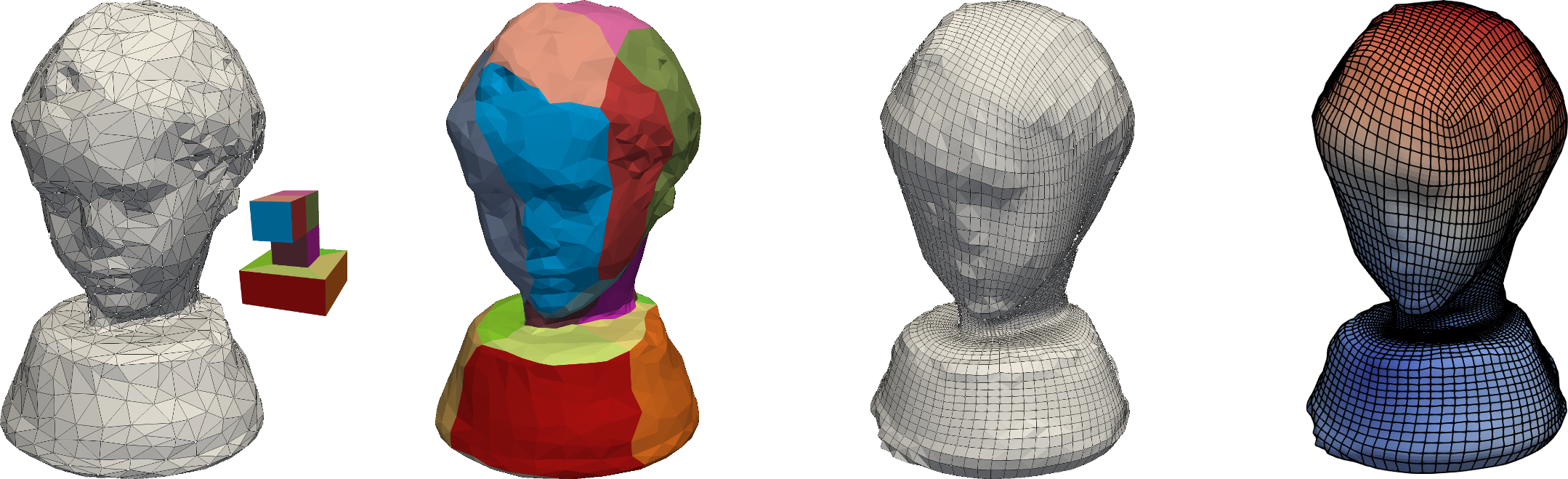}\\
  \begin{tikzpicture}
    \node[anchor=south west,inner sep=0] (image) at (0,0) {\includegraphics[width=0.8\linewidth]{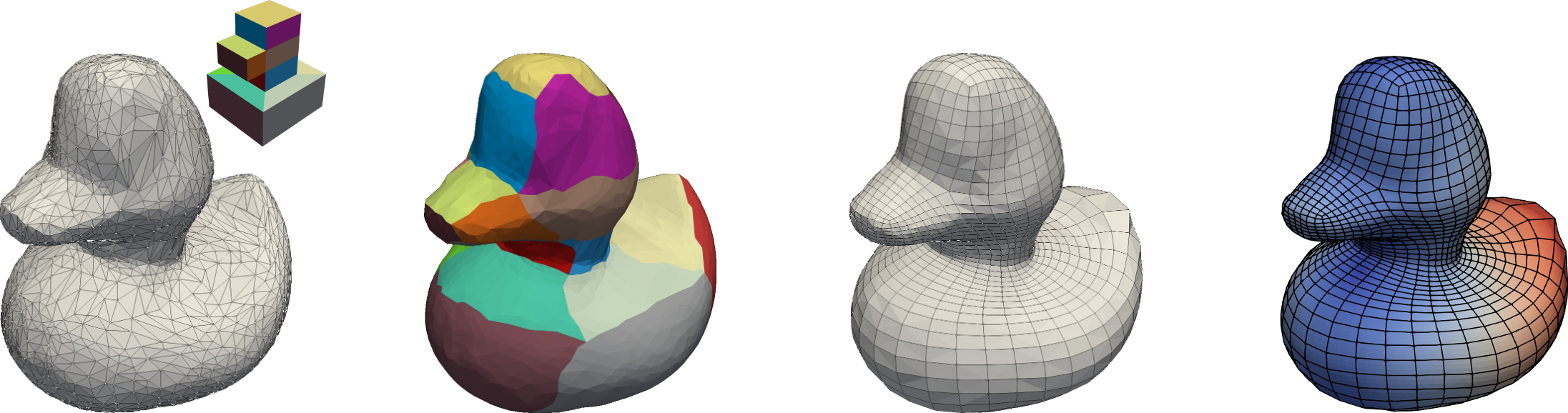}};
    \begin{scope}[x={(image.south east)},y={(image.north west)}]
      \node at (0.11,-0.1) {(a)};
      \node at (0.37,-0.1) {(b)};
      \node at (0.63,-0.1) {(c)};
      \node at (0.89,-0.1) {(d)};
    \end{scope}
  \end{tikzpicture}
\end{tabular}
\caption{ Results of CTP-cube, bust, and duck models. (a) Surface triangle meshes
  with its predicted polycube structures; (b) segmentation results; (c) all-hex
  control meshes; (d) volumetric splines with IGA eigenvalue analysis
  in ANSYS-DYNA. }
    \label{fig:model2}
\end{figure}
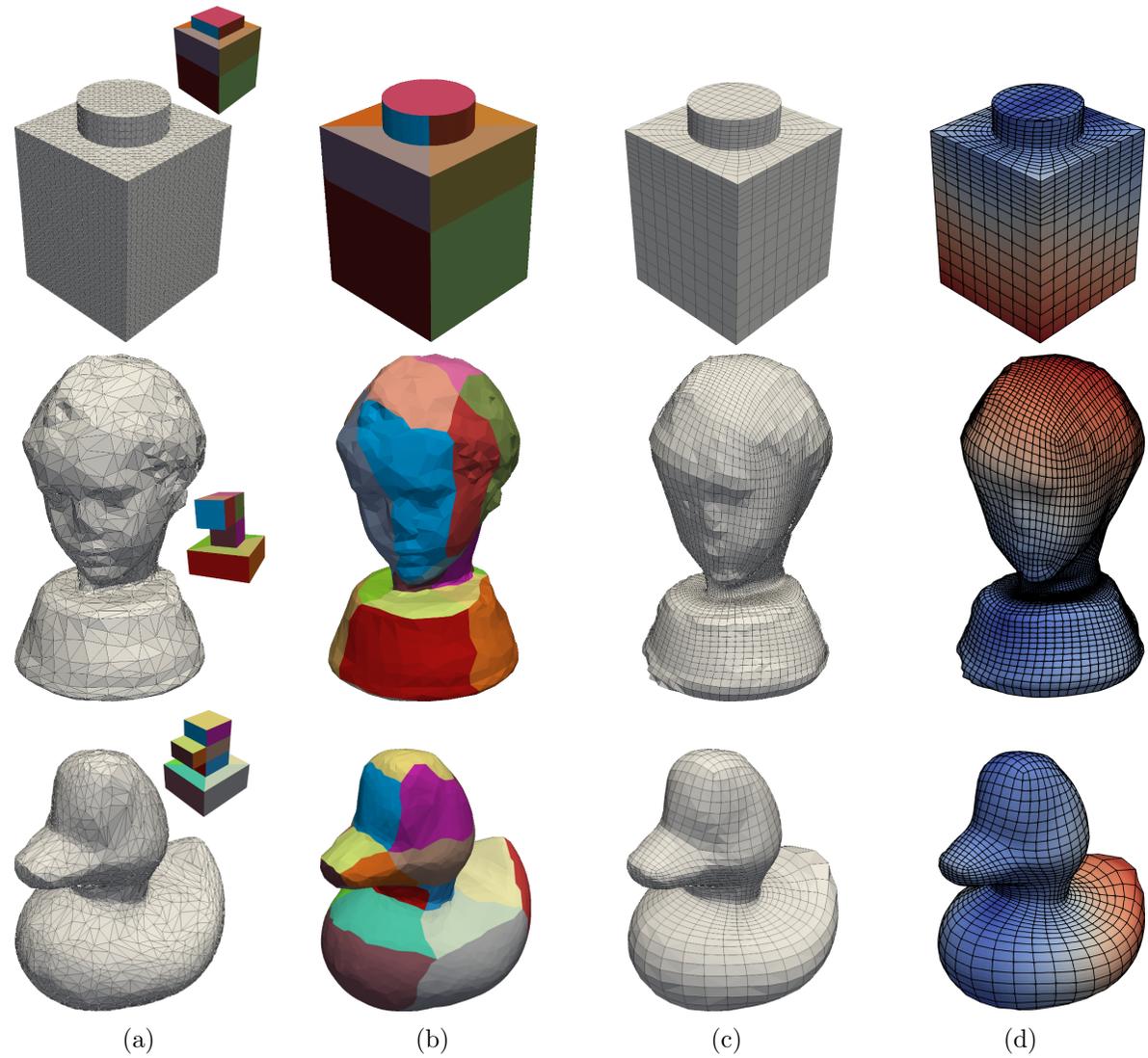
\end{landscape}

\begin{landscape}
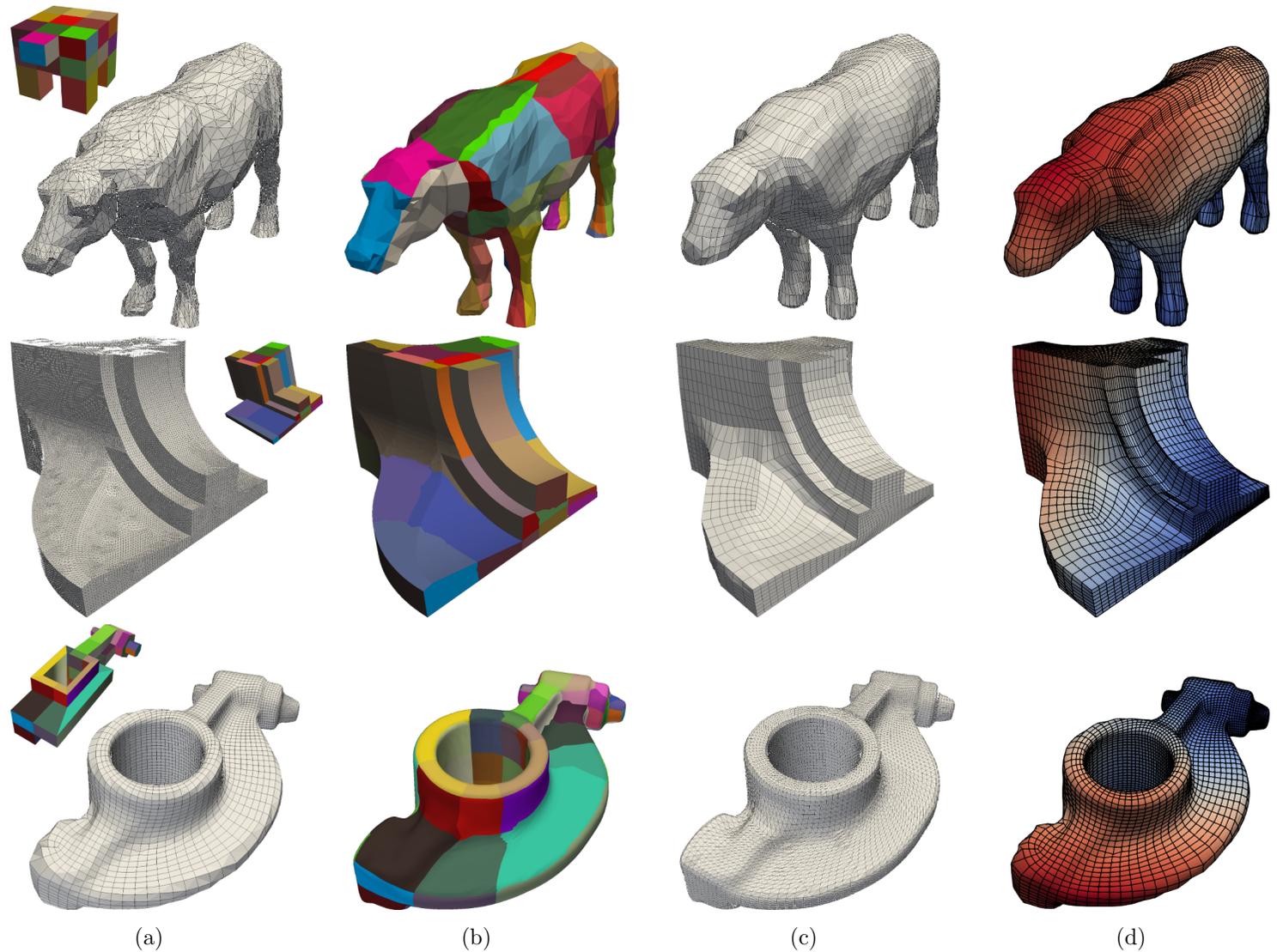
\begin{figure}[!htp]
\centering
\begin{tabular}{l}
  \includegraphics[width=\linewidth]{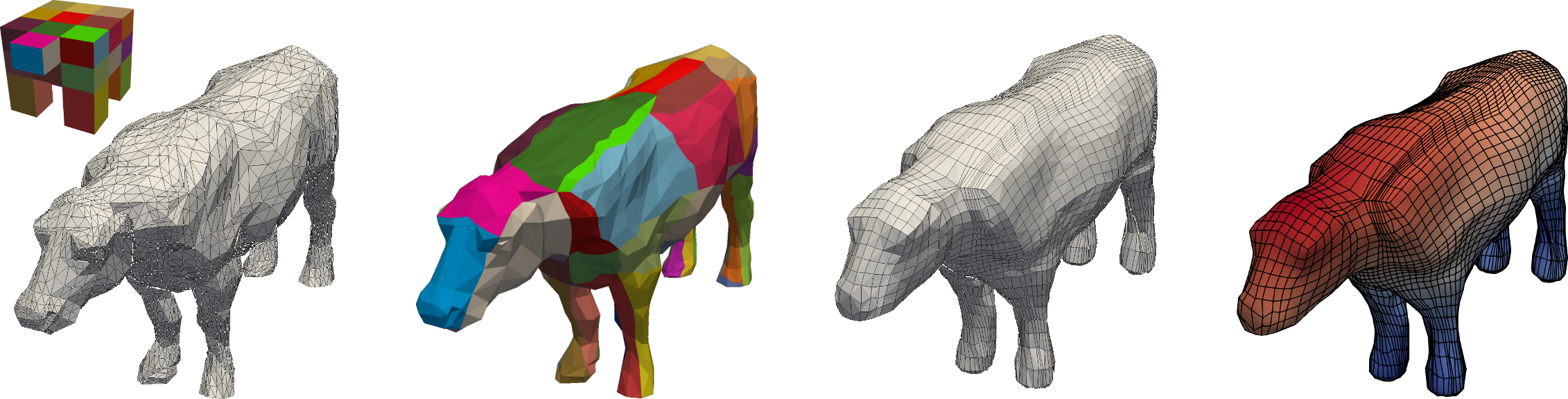}\\
  \includegraphics[width=\linewidth]{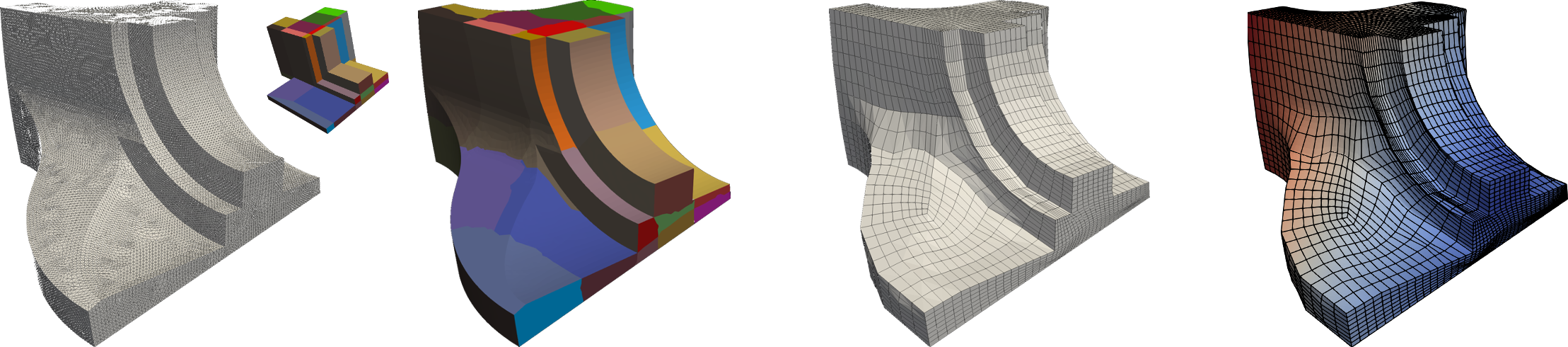}\\
  \begin{tikzpicture}
    \node[anchor=south west,inner sep=0] (image) at (0,0) {\includegraphics[width=\linewidth]{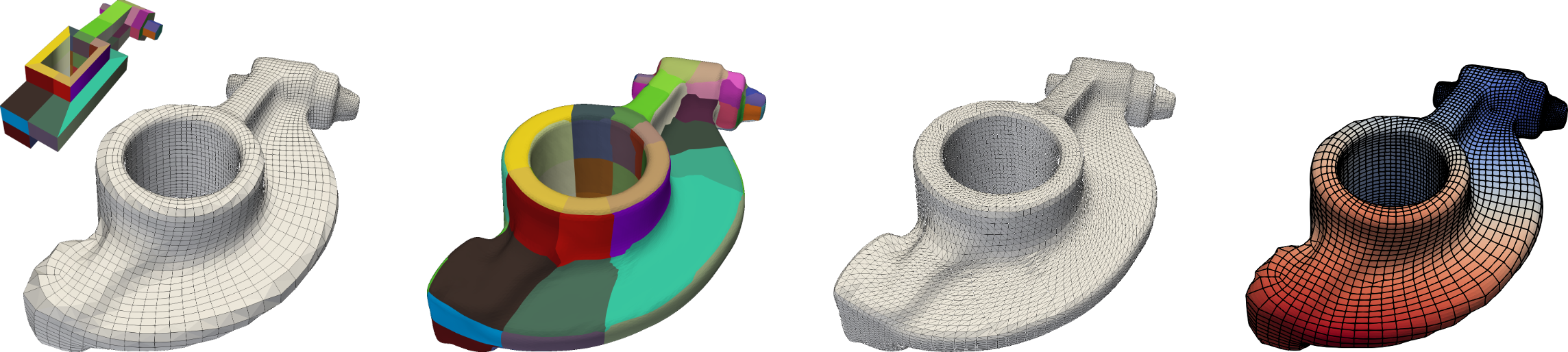}};
    \begin{scope}[x={(image.south east)},y={(image.north west)}]
      \node at (0.11,-0.1) {(a)};
      \node at (0.37,-0.1) {(b)};
      \node at (0.63,-0.1) {(c)};
      \node at (0.89,-0.1) {(d)};
    \end{scope}
  \end{tikzpicture}
\end{tabular}
\caption{ Results of cow, fandisk and rockerarm models. (a) Surface triangle meshes
  with its predicted polycube structures; (b) segmentation results; (c) all-hex
  control meshes; (d) volumetric splines with IGA eigenvalue analysis
  in ANSYS-DYNA. }
    \label{fig:model3}
\end{figure}
\end{landscape}

\bibliography{Hex-Software_edit}

\end{document}